\documentclass[superscriptaddress,amsmath,amssymb,nofootinbib,eqsecnum,a4paper,secnumarabic,11pt]{revtex4}         
\usepackage[pdftex]{graphicx}
\usepackage{hyperref}     
\usepackage{grffile} 
\usepackage{color}
\usepackage{braket}
\newcommand{\VEV}[1]{\langle #1 \rangle}
\usepackage{titlesec}     
\newcommand*{\justifyheading}{\raggedright}
\titleformat{\chapter}[display]
  {\normalfont\huge\bfseries\justifyheading}{\chaptertitlename\ \thechapter}
  {20pt}{\Huge}
\titleformat{\section}
  {\normalfont\Large\bfseries\justifyheading}{\thesection}{1em}{}
\titleformat{\subsection}
  {\normalfont\large\bfseries\justifyheading}{\thesubsection}{1em}{}
\titleformat{\subsubsection}
  {\normalfont\bfseries\justifyheading}{\thesubsubsection}{1em}{}
\usepackage[english]{babel}

\begin{document}
\begin{flushright}
{\tt 
OSU-HEP-19-10
}
\end{flushright}
\title{\vspace{-3truecm} \Large Simple theory of chiral fermion dark matter \vspace{5truemm}}
\author{\textbf{Tomohiro Abe}}
\email{abetomo@kmi.nagoya-u.ac.jp}
\affiliation{
  Institute for Advanced Research, Nagoya University,
  Furo-cho Chikusa-ku, Nagoya, Aichi, 464-8602 Japan
}
\affiliation{
  Kobayashi-Maskawa Institute for the Origin of Particles and the
  Universe, Nagoya University,
  Furo-cho Chikusa-ku, Nagoya, Aichi, 464-8602 Japan
}

\author{\textbf{K.S.~Babu}}
\email{babu@okstate.edu}
\affiliation{
  Department of Physics, Oklahoma State University, Stillwater, OKlahoma, 74078, USA
}

\vspace*{1.5in}

\begin{abstract}
\vspace{1truecm}
We propose a theory of chiral fermion dark matter (DM) with an isospin-3/2 fermion of a dark sector $SU(2)_D$ gauge symmetry, which is arguably the simplest chiral theory.  An isospin-3 scalar breaks $SU(2)_D$ down to a discrete non-Abelian group $T'$ and generates the DM mass. The $SU(2)_D$ gauge symmetry protects the DM mass and guarantees its stability. We derive consistency conditions for the theory and study its DM phenomenology. In some regions of parameters of the theory a two-component DM scenario is realized, consisting of a fermion and a boson, with the boson being the lightest $T'$ nonsinglet field.  In the case of single-component fermionic DM, we find that internal consistency of the theory, perturbativity arguments, and the observed relic abundance limit the DM mass to be less than $280$~GeV, except when $s$-channel resonance regions are open for annihilation. For a significant part of the parameter space, the theory can be tested in DM  direct detection signals at the LZ and XENONnT experiments.

\end{abstract}

\maketitle

\newpage

\section{Introduction}

The necessity for having a particle dark matter (DM) candidate is a compelling reason for physics beyond the Standard Model (SM).  There are indeed a variety of DM models in the literature. (For a review of weakly interacting massive particle (WIMP) dark matter candidates see Refs.~\cite{Cushman:2013zza, Arcadi:2017kky}.)  If the DM is thermally produced in the early Universe,
then its mass should be below ${\cal O}$(100)~TeV to explain the measured value of the DM energy density by the freeze-out mechanism.
This upper bound on the mass arises from unitarity constraints~\cite{Griest:1989wd}.
Why the DM mass lies in this range is an open question, with many DM models assuming this as an input. In the supersymmetric (SUSY) extension of the SM, the minimal supersymmetric SM, assuming $R$ parity, the lightest SUSY particle is a candidate for a thermal WIMP, and its mass scale may be justified if SUSY is to provide an explanation for the gauge hierarchy problem \cite{Roszkowski:2017nbc}.  The absence of any SUSY signature at the LHC thus far has prompted theoretical searches for well-motivated non-SUSY thermal DM candidates.  In this case a protection mechanism for its stability is needed, and it is highly desirable to have a symmetry-based mechanism to keep its mass 
below ${\cal O}$(100)~TeV.

Some of the proposed DM candidates, notably the ``minimal DM models''\cite{Cirelli:2005uq}, do provide a mechanism for the stability of DM based on the SM gauge symmetry.  For example, an $SU(2)_L$ quintuplet with zero hypercharge is absolutely stable and can explain the DM relic abundance if its mass is near $4.4$ TeV \cite{Cirelli:2009uv}.  However, such models do not explain why the DM mass is around $4.4$~TeV. With the DM candidate being a vector-like fermion under the SM gauge symmetry, its mass could have been much higher in principle. This comment also applies to minimal DM candidates which are elementary scalars.  (These minimal DM models are currently under stress from both direct detection searches~\cite{1608.07648, 1708.06917, 1805.12562} and indirect detection limits~\cite{Fermi-LAT:2016uux}.)

If the DM candidate is a chiral fermion under some gauge symmetry, its mass will be protected down to the symmetry-breaking scale, just as the top quark mass in the SM is protected down to the electroweak scale.  New fermions that are chiral under the electroweak symmetry  (e.g, a fourth-generation neutrino) are excluded as DM candidates, from both direct detection limits and from collider limits.  Thus the DM must be chiral under a new gauge symmetry.  In this paper, we propose what we believe is the simplest such possibility with the new gauge symmetry being $SU(2)_D$ and the DM fermion being the dark isospin-3/2 representation of this gauge symmetry.

The isospin-3/2 representation of $SU(2)_D$, a {\bf 4}-plet, is the simplest chiral representation of the group that is anomaly free.  Chiral theories are constrained by the requirement that they should be free of all anomalies. While the gauge group $SU(2)$ is devoid of triangle anomalies, fermion representations are subject to the global Witten anomaly~\cite{Witten:1982fp} which is determined by the quadratic index $\mu$ of the fermion representations.  For representation $I$ the quadratic index is given by $\mu = 2I(I+1)(2I+1)/3$, from which it follows that for $I=3/2$, $\mu=10$, which is an even integer.  Representations with even $\mu$ have no global anomaly and are consistent gauge theories, while those with odd $\mu$ suffer from the Witten anomaly and are inconsistent.  For example, an isospin-1/2 fermion will have $\mu = 1$ which is anomalous.  $SU(2)$ theories with isospin-1/2 fermions must have them in even numbers, in which case bare mass terms can be written for all of them, making such theories vector-like. Similarly, a triplet fermion (with $\mu = 4$) is allowed to have a bare mass term and is not chiral. Models with an isospin-3/2 fermion are the simplest choice that is anomaly free and chiral in the sense that one cannot write a bare mass term for it. Note that the singlet state obtained from $\mathbf{4} \otimes \mathbf{4}$ of $SU(2)$ is antisymmetric,
which makes its mass term forbidden.
Models based on $SU(2)$ gauge symmetry with an isospin-3/2 fermion have been studied in the context of dynamical supersymmetry breaking~\cite{hep-ph/9410203} which require chiral theories, as well as for realizing light sterile neutrinos~\cite{hep-ph/0312285}.  Here we develop such a theory to explain the DM content of the Universe. The chiral fermion will acquire its mass only after spontaneous symmetry breaking by the Higgs mechanism. The scale of this symmetry breaking will be assumed to be around the TeV scale, analogous to the SM Higgs mechanism, where the scale of symmetry breaking is near $100$~GeV.

It is clear that any representation of $SU(2)$ with $ I = 3/2+2n$ (where $n$ is a non-negative integer) will have an even index, and thus will lead to a chiral theory.  Thus, the next simplest case would be a theory with an $I = 7/2$ fermion with an index of $\mu = 84$.  However, in this case the theory would lose asymptotic freedom, with the beta function coefficient given by $b_D = 62/3$ (where $(16 \pi^2) dg_D/dt = b_D g_d^3$). On the other hand, with the $I=3/2$ theory, $b_D = -4$, which implies that this theory can be extrapolated to high energies. It should be noted that such chiral theories are somewhat special to $SU(2) \approx Sp(1)$ and to our knowledge have no generalizations to arbitrary $SU(n)$.

It is worth noting that anomaly free chiral gauge theory based on $U(1)$, which may appear to be simpler, requires at least five fermion fields \cite{hep-ph/0312285,Sayre:2005yh,Batra:2005rh,Costa:2019zzy,Berryman:2016rot} and is more elaborate.  Chiral fermionic DM has been studied in the context of a new $SU(3)' \times SU(2)'$ gauge symmetry with fermion representations that are parallel to one family of SM fermions in Ref. \cite{Berryman:2017twh}.  We believe that the model of $SU(2)_D$ presented here is much simpler.  

We have developed a theory based on $SU(2)_D$ gauge symmetry and an isospin-3/2 fermionic representation serving as the DM candidate. The simplest way to generate mass for the fermion and the gauge bosons is by introducing an isospin-3 Higgs multiplet, a {\bf 7}-plet.  Since this theory is quite simple with very few parameters, it is rather constrained theoretically as well as from DM detection experiments which we outline.  Partial wave unitarity in gauge boson scattering to scalar bosons sets an upper limit on the $SU(2)_D$ gauge coupling $g_D < 0.985.$  Fermion-fermion scattering to two scalar bosons limits the single Yukawa coupling of the model to be $|y_D| < 2.4.$ Two-to-two scattering in the scalar sector also sets stringent limits on the quartic scalar couplings of the theory.  When combined, these internal consistency conditions lead to a predictive DM scenario.  In the single component fermionic scenario, we find the DM mass should be less than $\sim 280$~GeV, except in regions of resonant enhancement in the $s$-channel for the annihilation cross section.  Since the isospin-3 Higgs multiplet leaves a non-Abelian discrete symmetry $T'$ intact, for some range of mass parameters the model predicts a two-component DM scenario with the fermion as well as the lightest $T'$-nonsinglet boson contributing to the relic abundance.  We map out the regions of parameters consistent with this scenario.

The rest of this paper is organized as follows.  In Sec.~\ref{sec:Model}, we describe the chiral DM theory setup. Various theoretical constraints on the parameters of the theory are derived here.  In Sec.~\ref{sec:pheno}, we study the DM phenomenology of the theory.
Here we show that the single component fermionic DM should be lighter than $\sim 280$~GeV unless its annihilation relies  on $h'$ resonance, where $h'$ is a singlet scalar, a remnant of the $SU(2)_D$ symmetry breaking. We also map out the parameter space of the theory with two-component DM in this section.   Section~\ref{sec:conclusion} is devoted to our conclusions.

\section{\boldmath{$SU(2)_D$} theory of chiral fermion DM}\label{sec:Model}

As motivated in the Introduction, we extend the SM by a dark sector gauge symmetry $SU(2)_D$.  The fermion content is very simple, with an isospin-3/2 multiplet $\psi$, which is an SM singlet. The Higgs sector is also simple, with an isospin-3 multiplet $\phi$, which is also an SM singlet.  These fields are parametrized as:
\begin{align}
\psi^t
=&
\begin{pmatrix}
 \psi_{3/2},\ \psi_{1/2},\ \psi_{-1/2},\ \psi_{-3/2}
\label{eq:4-plet-for-SU2}
\end{pmatrix}
,\\
\phi^t
=&
 \begin{pmatrix}
  \phi_3,\   \phi_2,\   \phi_1,\   \phi_0,\ \phi_{-1},\   \phi_{-2},\   \phi_{-3} 
 \end{pmatrix}
.
\end{align}
Because the 7-plet is a self-dual field, its components are subject to the following relations:
\begin{align}
 \phi_{-3} = - \phi_3^*,\quad
 \phi_{-2} = + \phi_2^*,\quad
 \phi_{-1} = - \phi_1^*,\quad
 \phi_{0} =  + \phi_0^*.
\end{align}
The Higgs potential for $\phi$ (and the SM doublet $\Phi$) admits a minimum where 
$\VEV{\text{Im}\phi_2} \equiv -\frac{v_D}{\sqrt{2}} \neq 0$, 
$\VEV{\text{Re}\phi_2} \equiv 0$, 
and $\VEV{\phi_j} = 0$ for $j \neq 2$.
In this vacuum, $SU(2)_D$ is broken down to $T'$ which is the double covering group of $A_4$~\cite{Koca:1997td,hep-ph/0410270,Berger:2009tt}. In the absence of half-integer representations, a 7-plet would have broken $SU(2)_D$ down to $A_4$, but in our case the full theory has a fermionic 4-plet, which implies that the surviving symmetry is $T'$. This unbroken $T'$ plays a crucial role in DM phenomenology.  Under $T'$, a 4-plet decomposes as ${\bf 4} = {\bf 2'} + {\bf 2''}$, while ${\bf 7} = {\bf 1} + {\bf 3} + {\bf 3}$. The vacuum expectation value (VEV) of $\phi$  generates equal mass for the three gauge bosons, which transform as a {\bf 3} of $T'$.  The Higgs sector will consist of a physical {\bf 3}-plet, a Goldstone {\bf 3}-plet and a $T'$ singlet $h'$ (apart from the SM Higgs boson $\Phi$, which is also a $T'$ singlet).  

One could have considered achieving $SU(2)_D$ symmetry breaking via other Higgs representations.  It is economical if this Higgs also provides the 4-plet fermion with a mass.  The choice for the Higgs is then narrowed down to either a 3-plet or 7-plet.  A Higgs triplet would leave an unbroken $U(1)$, which would result in dark radiation in the early Universe, which is excluded by big bang nucleosynthesis (BBN) and highly constrained by Planck data~\cite{Aghanim:2018eyx}.  Using two copies of a triplet Higgs is possible, but that would introduce more parameters than the use of a single 7-plet of Higgses.

The stability of the DM candidate in the theory can be understood as follows.  Since there is no mixing between $\psi$ and the SM fermions, fermion number conservation would imply that $\psi$ is stable.  Furthermore, the lightest nonsinglet particle of $T'$ should be stable, as $T'$ is an unbroken discrete gauge symmetry.  The fermions from the 4-plet may be the lightest $T'$ nonsinglet fields, as they transform as ${\bf 2'} + {\bf 2''}$ under $T'$. In this case the fermions which have a common mass is the only DM candidate.  It is also possible that the fermion is heavier than either the gauge boson multiplet, or the physical Higgs multiplet, both of which transform as {\bf 3} of $T'$.  In this case we realize a two-component DM, with a fermion and a boson. We shall analyze the DM phenomenology of both scenarios.

We note parenthetically that a 7-plet of Higgses could also break $SU(2)_D$ down to $Q_6$, the double cover of $D_3$~\cite{Koca:1997td, hep-ph/0410270,Berger:2009tt}.  However, if we adopt this breaking chain, we found that two of the components of the fermions would remain massless, which is excluded by BBN and Planck data~\cite{Aghanim:2018eyx}.  Thus, we restrict to our scenario of $SU(2)_D \rightarrow T'.$

The Yukawa interaction terms of the dark sector in the theory are given by
\begin{align}
 {\cal L}_{Yukawa}
 =&
 - y_D
\Biggl[
 \phi_{-3} \psi_{3/2} \psi_{3/2}
- \sqrt{2} \phi_{-2} \psi_{1/2} \psi_{3/2}
+ \phi_{-1}\left( \sqrt{\frac{3}{5}} \psi_{1/2} \psi_{1/2} + \frac{2}{\sqrt{5}} \psi_{3/2} \psi_{-1/2}  \right)
\nonumber\\
& \qquad \quad
- \phi_0 \left( \frac{3}{\sqrt{5}} \psi_{1/2} \psi_{-1/2} + \frac{1}{\sqrt{5}} \psi_{3/2} \psi_{-3/2} \right)
\nonumber\\
& \qquad \quad
+ \phi_1 \left( \sqrt{\frac{3}{5}} \psi_{-1/2} \psi_{-1/2} + \frac{2}{\sqrt{5}} \psi_{-3/2} \psi_{1/2}  \right)
- \sqrt{2} \phi_{2} \psi_{-1/2} \psi_{-3/2}
+\phi_{3} \psi_{-3/2} \psi_{-3/2}
\Biggr]
\nonumber\\
&+(H.c.)
.
\label{Yuk}
\end{align}
The coefficients here are determined by the standard Clebsch-Gordan decomposition.
A bare mass term for the fermion is forbidden (since $\mathbf{4} \otimes \mathbf{4}$ yielding a singlet is antisymmetric) and thus $\psi$ acquires its mass from the Yukawa interaction terms after the 7-plet scalar field develops its VEV. 
The scalar potential of the model is given by
\begin{align}
 V_{scalar} 
=&
\mu_H^2 \Phi^\dagger \Phi 
+ \lambda (\Phi^\dagger \Phi)^2 
+ m^2 \left( \frac{1}{2} \phi_0^2 + |\phi_1|^2 + |\phi_2|^2  + |\phi_3|^2 \right)
\nonumber\\
&
+ \lambda_{H\phi} \Phi^\dagger \Phi \left( \phi_0^2 + 2 |\phi_1|^2 + 2 |\phi_2|^2  + 2 |\phi_3|^2 \right)
+ \lambda_1 \left( \frac{1}{2} \phi_0^2 + |\phi_1|^2 + |\phi_2|^2  + |\phi_3|^2 \right)^2
\nonumber\\
&
 + \lambda_2 
\Biggl(
\left| \sqrt{6} \phi_1^2 - 2 \sqrt{5} \phi_0 \phi_2 + \sqrt{10} \phi_3 \phi_{-1} \right|^2
+ 
\left| \sqrt{2} \phi_0 \phi_1 - \sqrt{15} \phi_2 \phi_{-1} + 5 \phi_3 \phi_{-2} \right|^2
\nonumber\\
&
\qquad \quad
+ \frac{1}{2}
\left( 2 \phi_0^2 + 3 |\phi_1|^2  - 5 |\phi_3|^2 \right)^2
\Biggr)
,
\end{align}
where $\Phi$ is the SM Higgs field.
The construction of the $\lambda_1$ term is trivial.
We construct the $\lambda_2$ term from the 5-plet obtained from two 7-plet fields,
$
\left(
(\phi \otimes \phi)_\mathbf{5} \otimes (\phi \otimes \phi)_\mathbf{5}
\right)_\mathbf{1}
$.
We can also construct other quartic terms, 
but it can be shown that they are linear combinations of the $\lambda_1$ and $\lambda_2$ terms.

We assume the vacuum structure of the Higgs fields to be
$\VEV{\text{Im}\phi_2} \equiv -\frac{v_D}{\sqrt{2}} \neq 0$, 
$\VEV{\text{Re}\phi_2} \equiv 0$, 
and $\VEV{\phi_j} = 0$ for $j \neq 2$.
We also assume that $\Phi$ breaks $SU(2)_L \times U(1)_Y$,
$\VEV{\Phi} = (0, \frac{v}{\sqrt{2}})^t$.
The mass parameters in the scalar potential and the VEVs are related by
\begin{align}
 \mu_H^2 =& -\lambda v^2 - v_D^2 \lambda_{H\Phi},\\
m^2 =& - v_D^2 \lambda_1  - v^2 \lambda_{H\Phi}.
\end{align}
In this vacuum, $SU(2)_D$ is broken down to $T'$, which is the double covering group of $A_4$~\cite{Koca:1997td, hep-ph/0410270}. 
The 7-plet scalar decomposes as 
\textbf{3}$+$\textbf{3}$+$\textbf{1} under $T'$ or $A_4$.

After the Higgs fields develop VEVs, 
the scalar sector contains a massless $T'$-triplet, a massive $T'$-triplet, and two massive $T'$-singlet fields.
The three massless fields are would-be Nambu-Goldstone (NG) bosons that are eaten by the $SU(2)_D$ gauge bosons.
The mass eigenstates of the triplet fields are given by
\begin{align}
 \pi_1 =& \frac{\sqrt{3}}{2} \text{Re} \phi_3 + \frac{\sqrt{5}}{2} \text{Re} \phi_1,\\
 \pi_2 =& -\frac{\sqrt{3}}{2} \text{Im} \phi_3 + \frac{\sqrt{5}}{2} \text{Im} \phi_1,\\
 \pi_3 =& \sqrt{2}\text{Re}\phi_2,\\
 H_1 =& -\frac{\sqrt{5}}{2} \text{Re} \phi_3 + \frac{\sqrt{3}}{2} \text{Re} \phi_1,\\
 H_2 =& -\frac{\sqrt{5}}{2} \text{Im} \phi_3 - \frac{\sqrt{3}}{2} \text{Im} \phi_1,\\
 H_3 =& \phi_0,
\end{align}
where $\pi_j$ are the would-be NG bosons.
The squared mass of the $H_j$ triplet is given by
\begin{align}
 m^2_{H} =& 20 \lambda_2  v_D^2.
\end{align}
We define the $\sigma_7$ field as
\begin{align}
 \text{Im} \phi_2 = -\frac{v_D + \sigma_7}{\sqrt{2}}.
\end{align}
The mass terms for the $T'$-singlet scalar fields are given by
\begin{align}
 -\frac{1}{2}
  \begin{pmatrix} 
    \sigma & \sigma_7 
  \end{pmatrix}
  \begin{pmatrix}
   2 \lambda v^2 &  2\lambda_{H\phi} v v_D \\
   2 \lambda_{H\phi} v v_D &  2 \lambda_1  v_D^2 
\end{pmatrix}
 \begin{pmatrix} 
     \sigma \\ \sigma_7 
   \end{pmatrix}
,
\end{align}
where $\sigma$ is the CP-even scalar field in the SM Higgs doublet.
We define $h$, $h'$, and $\theta$ by
\begin{align}
 \begin{pmatrix}
  h \\ h'
 \end{pmatrix} 
=
\begin{pmatrix}
 \cos\theta & -\sin\theta \\
\sin\theta & \cos\theta
\end{pmatrix}
 \begin{pmatrix}
  \sigma \\ \sigma_7
 \end{pmatrix} 
.
\end{align}
The masses of the $h$ and $h'$ are related to the mass matrix by
\begin{align}
 \begin{pmatrix}
   2 \lambda v^2 &  2 \lambda_{H\Phi} v v_D \\
   2 \lambda_{H\Phi} v v_D &  2 \lambda_1 v_D^2 
\end{pmatrix}
=
\begin{pmatrix}
 \cos\theta & \sin\theta \\
-\sin\theta & \cos\theta
\end{pmatrix}
\begin{pmatrix}
 m_h^2 & 0 \\ 0 & m_{h'}^2 
\end{pmatrix}
\begin{pmatrix}
 \cos\theta & -\sin\theta \\
\sin\theta & \cos\theta
\end{pmatrix}
.
\end{align}

The three SU(2)$_D$ gauge bosons, which form a $T'$ triplet, acquire a common mass from the 7-plet VEV given by
\begin{align}
 m_{V}^2 =& 4 g_D^2 v_D^2,
\end{align}
where $g_D$ is the SU(2)$_D$ gauge coupling.

The 4-plet fermion acquires its mass from the Yukawa interaction terms of Eq. (\ref{Yuk}) as
\begin{align}
 {\cal L}\supset i y_D v_D \left(\psi_{3/2} \psi_{1/2} - \psi_{-1/2} \psi_{-3/2}\right).
\end{align}
As can be seen, there are two Dirac fermions after the SU(2)$_D$ symmetry breaking. 
Their masses are equal and are given by
\begin{align}
 m_\Psi = y_D v_D.
\end{align}
These Dirac fermions are $T'$ doublets. 
Note that there is no four-dimensional irreducible representations in $T'$, 
and thus the 4-plet fermion field is decomposed into irreducible representations as $\mathbf{4} = \mathbf{2'} + \mathbf{2''}$.
We find that the following chiral fermions are doublets of $T'$:
\begin{align}
 \psi_{\mathbf{2''}} =& \begin{pmatrix}
			 \xi^1 \\ \xi^2
			\end{pmatrix}, \quad
 \psi_{\mathbf{2'}} = \begin{pmatrix}
			 \chi^1 \\ \chi^2
			\end{pmatrix}, \label{eq:2-of-T'}
\end{align}
where
\begin{align}
 \xi^1 =& \frac{i \psi_{1/2} +  \psi_{-3/2}}{\sqrt{2}}, \label{eq:xi1}\\
 \xi^2 =& \frac{-\psi_{3/2} - i \psi_{-1/2}}{\sqrt{2}}, \label{eq:xi2}\\
 \chi^1 =& \frac{-i \psi_{1/2} + \psi_{-3/2}}{\sqrt{2}}, \label{eq:chi1}\\
 \chi^2 =& \frac{-\psi_{3/2} + i \psi_{-1/2}}{\sqrt{2}}. \label{eq:chi2}
\end{align}
The Dirac fermion and its charge conjugate are constructed from $\psi_{\mathbf{2'}}$ and $\psi_{\mathbf{2''}}$ as
\begin{align}
 \Psi^j 
=&
 \begin{pmatrix}
 (\psi_\mathbf{2''})^j \\ (i \sigma_2 \psi^\dagger_{\mathbf{2'}})^j
 \end{pmatrix}
,\quad
 (\Psi^c)^j 
= 
 \begin{pmatrix} 
(\psi_\mathbf{2'})^j \\ (i \sigma_2 \psi^\dagger_\mathbf{2''})^j 
\end{pmatrix},
\end{align}
where $j$ is the index for $T'$.
We find that $\Psi$ and $\Psi^c$ transform as $\mathbf{2''}$ and $\mathbf{2'}$, respectively; see 
Appendix~\ref{sec:T'rules} for more details.
The charge conjugate of $\Psi$ is needed to write down all of the interaction terms as
shown in Appendix~\ref{app:fermion-int}.

\subsection{Parameters of the theory}

There are eight parameters in the dark sector plus the SM scalar sector:
\begin{align}
 \left(
  \mu_H^2, \lambda,  \lambda_{H\phi}, m^2, \lambda_1, \lambda_2, g_D, y_D
 \right).
\end{align}
Instead of these parameters, we use the followings as input:
\begin{align}
 \left(
  v, m_h,  \theta, m_{\Psi}, m_{h'}, m_{H}, m_V, y_D
 \right).
\end{align}
Using these input parameters, the other parameters can be written as follows:
\begin{align}
 v_D =&  \frac{m_\Psi}{y_D}, \\
 g_D =&  \frac{m_V}{2 m_\Psi} y_D, \label{eq:gD}\\
 \lambda =& \frac{m_h^2 \cos^2\theta + m_{h'}^2  \sin^2\theta}{2 v^2}, \label{eq:lam}\\
 \lambda_1 =& \frac{m_{h'}^2 \cos^2\theta + m_h^2 \sin^2\theta}{2 m_\Psi^2} y_D^2, \label{eq:lam1}\\
 \lambda_2 =& \frac{m_H^2}{20 m_\Psi^2} y_D^2, \label{eq:lam2}\\
 \lambda_{H\phi} =& \frac{(m_{h'}^2-m_h^2) \cos\theta \sin\theta}{2 v m_\Psi } y_D. \label{eq:lamHPhi}
\end{align}
It is important to know the region of parameter space where the couplings are within the perturbative regime.
We assume that the Higgs mixing angle $\theta$ is small 
in order to avoid LHC constraints on the Higgs coupling strengths as well as limits from DM direct detection experiments.
For small $\theta$, $\lambda_{H\Phi}$ is suppressed by $\theta$ and its absolute value is much smaller than unity in
most regions of the parameter space.
Similarly, $\lambda$ has almost the same value as the Higgs quartic coupling in the SM for small $\theta$.
The other parameters---$\lambda_1$, $\lambda_2$, and $g_D$---can take values larger than unity if the mass differences of the particles in the dark sector are large. We turn this argument around and constrain the mass differences by requiring perturbative values of these couplings.

\subsection{Perturbative unitarity constraints}\label{sec:PU}

We first focus on perturbative unitarity limits on the quartic scalar couplings of the theory.  
The High energy two-to-two scattering amplitude can be expressed as
\begin{align}
 {\cal M} = 16 \pi \sum_J (2J +1) a_J P_J(\cos\theta).
\end{align}
We consider the processes that only contain the 7-plet field in both initial and final states.
There are 28 combinations of the states for the initial and final states, given by,
\begin{align}
& \frac{1}{\sqrt{2}} \ket{\sigma_7 \sigma_7},\
 \frac{1}{\sqrt{2}} \ket{H_1 H_1},\
 \frac{1}{\sqrt{2}} \ket{H_2 H_2},\
 \frac{1}{\sqrt{2}} \ket{H_3 H_3},\
 \frac{1}{\sqrt{2}} \ket{\pi_1 \pi_1},\
 \frac{1}{\sqrt{2}} \ket{\pi_2 \pi_2},\
 \frac{1}{\sqrt{2}} \ket{\pi_3 \pi_3},\\
&\ket{\sigma_7 H_1},\
 \ket{\sigma_7 H_2},\
 \ket{\sigma_7 H_3},\
 \ket{\sigma_7 \pi_1},\
 \ket{\sigma_7 \pi_2},\
 \ket{\sigma_7 \pi_3},\\
& \ket{H_1 H_2},\
 \ket{H_1 H_3},\
 \ket{H_2 H_3},\
 \ket{\pi_1 \pi_2},\
 \ket{\pi_1 \pi_3},\
 \ket{\pi_2 \pi_3},\\
& \ket{H_1 \pi_1},\
 \ket{H_1 \pi_2},\
 \ket{H_1 \pi_3},\
 \ket{H_2 \pi_1},\
 \ket{H_2 \pi_2},\
 \ket{H_2 \pi_3},\
 \ket{H_3 \pi_1},\
 \ket{H_3 \pi_2},\
 \ket{H_3 \pi_3}.
\end{align}
Therefore, the $s$-wave amplitude $a_0$ is a 28 $\times$ 28 matrix.
We calculate the eigenvalues of $a_0$ and use the maximum eigenvalue, $|\text{Re}(a_0^{\text{max}})|$, to derive a constraint on $\lambda_1$ and $\lambda_2$ from perturbative unitarity (PU)~\cite{Lee:1977eg}:
\begin{align}
|\text{Re}(a_0^{\text{max}})| < \frac{1}{2}.  
\end{align}
We find four independent eigenvalues, and the PU bound is given by
\begin{align}
 8 \pi > \max\left\{ \left|2\lambda_1 - 30\lambda_2 \right|, \
 \left|2 \lambda_1 + 25 \lambda_2 \right|, \
 \left|9\lambda_1 + 60\lambda_2 \right|, \
 \left|2 \lambda_1 + 61\lambda_2 \right| \right\}.
\label{eq:PU}
\end{align}
The third eigenvalue is the largest in most regions of the parameter space due to the large numerical coefficients. The corresponding eigenvector is given by
\begin{align}
\frac{1}{\sqrt{14}}
\biggl(
 \ket{\sigma_7 \sigma_7}
 + \ket{H_1 H_1}
 + \ket{H_2 H_2}
 + \ket{H_3 H_3}
 + \ket{\pi_1 \pi_1}
 + \ket{\pi_2 \pi_2}
 + \ket{\pi_3 \pi_3}
\biggr).
\end{align}

Now we turn to the PU limit on the $SU(2)_D$ gauge coupling $g_D$.  It is possible to obtain such a bound from the $\phi \phi \to VV$ scattering channel, where $V$ is a transversely polarized vector boson. 
If $\phi$ is an $SU(2)$ real $n$-plet scalar field, the PU bound on the gauge coupling is given by~\cite{1202.5073}
\begin{align}
g_D^2 \frac{(n^2 - 1)\sqrt{n}}{2 \sqrt{6}} < 8 \pi.
\end{align}
Applying this result to $n=7$, we find
\begin{align}
 |g_D| < 0.985.
\end{align}
This is a stringent bound for the single component fermionic DM, as we discuss later.

We also study $\Psi^i_h \bar{\Psi}^j_{h'} \to \pi^a_V \pi^b_V$ and 
$\Psi^i_h \bar{\Psi}^j_{h'} \to H_a H_b$ processes, where $i$ and $j$ are the indices for $T'$, 
and $h$ and $h'$ are twice the helicity of each fermion. 
We consider the $s$-wave amplitude in the high energy limit and take the $T'$-singlet state for the initial and final states, which is proportional to $\frac{1}{\sqrt{2}} \delta^{ij} \frac{1}{\sqrt{3}} \delta^{ab}$. 
We find that $\Psi \bar{\Psi} \to \pi^a_V \pi^b_V$ does not contain the singlet state in the high energy limit.
We also find that ${\cal M}_{i+,j+}^{ab}$ and ${\cal M}_{i-,j-}^{ab}$ are suppressed by ${\cal O}(m/\sqrt{s})$ in the high energy limit. For the other helicity combinations, we find       
\begin{align}
 (a_0)_{+,-}
=(a_0)_{-,+}
= \frac{1}{16\pi} \frac{4\pi}{5} y_D^2 \sqrt{3} \sqrt{2}.
\end{align}
Since the final state contains identical particles, we impose $|a_0| < \frac{1}{\sqrt{2}}$ and obtain the limit
\begin{align}
 |y_D| < \left( \frac{10}{\sqrt{3}} \right)^{1/2} \simeq 2.4.
\end{align}

\subsection{Boundedness of the potential}

The scalar potential should be bounded from below.
Here we discuss the boundedness conditions of the scalar potential in the large field value regime and give constraints on the quartic couplings. 

It is easy to see that $\lambda$ should be positive by checking the boundedness of the potential for $\vec{\phi} = \vec{0}$.
It is also easy to find the condition on $\lambda_1$
because the $\lambda_2$ term has flat directions. For example,
the $\lambda_2$ term vanishes for
$\phi_1 = \phi_2 = 0$ and $|\phi_3| = \sqrt{\frac{2}{5}} \phi_0$.
Along this direction with $\Phi = 0$, the $\lambda_1$ should be positive for the potential to be bounded from below.
Other necessary and sufficient conditions are obtained after some algebra. 
We find these conditions to be\footnote{
Our result is consistent with the result of  Ref.~\cite{1510.01559}.  The following translation of notation is needed:
$
 \lambda_2 + \lambda_4 \to \lambda_1,
 \lambda_3 \to 2 \lambda_{H\phi},
 \lambda_4 \to 16 \lambda_2.
$
}
\begin{align}
    \lambda >& 0,\\
    \lambda_1 >& 0,\\
    \tilde{\lambda}_1>& 0,\\
    \lambda_{H\phi} + \sqrt{\lambda \tilde{\lambda}_1} >& 0, \label{eq:bfb-last}
\end{align}
where
\begin{align}
   \tilde{\lambda}_1 = \lambda_1 + \min\left(0, \frac{25}{2} \lambda_2 \right).
\end{align}

We can express these four conditions by physical observables by using Eqs.~\eqref{eq:lam}--\eqref{eq:lamHPhi}.
As shown in Eqs.~\eqref{eq:lam}--\eqref{eq:lamHPhi}, 
$\lambda$, $\lambda_1$, and $\lambda_2$ are positive,
and thus the first three conditions are automatically satisfied.
The last condition can be expressed as
\begin{align}
    (m_{h'}^2 - m_h^2) \cos\theta \sin\theta
    + 
    \sqrt{(m_{h'}^2 - m_h^2)^2 \cos^2\theta \sin^2\theta + m_h^2 m_{h'}^2}
    >0.
\end{align}
This inequality is always satisfied in our minimum. 
Therefore, the scalar potential is always bounded from below and poses no new conditions.

\subsection{Higgs coupling strength}
Through the scalar mixing, the Higgs couplings to the SM particles are influenced
and are scaled by $\cos\theta$. Namely,
\begin{align}
    \frac{g_{WWh}}{g_{WWh}^\text{SM}} 
    =\frac{g_{ZZh}}{g_{ZZh}^\text{SM}} 
    =\frac{g_{ffh}}{g_{ffh}^\text{SM}} 
    = \cos\theta ,
\end{align}
where $f$ stands for the SM fermions. 
The measurements of the Higgs couplings give the upper bound on the mixing angle $\theta$.
The ATLAS experiment~\cite{Aad:2019mbh} found the following constraints on these ratios of the couplings to the SM predictions, 
\begin{align}
    \kappa_V =& 1.05 \pm 0.04,\\
    \kappa_F =& 1.05 \pm 0.09.
\end{align}
The observed linear correlation between them is 44\%.
Using this result, we find that
\begin{align}
    |\theta| < 0.27  \quad \text{(90\% C.L.)}.
\end{align}

\subsection{Direct detection}
In our model, $\Psi$ is always a dark matter particle.
The lighter $T'$-triplet field is also a dark matter candidate if its decay is kinematically forbidden. Thus we show the cross section of a $T'$ nonsinglet elastically scattering off of a nucleon.
\begin{align}
    \sigma_\text{SI} (\Psi N \to \Psi N)
    =& s_\theta^2 c_\theta^2 
    \left( \frac{1}{m_h^2} - \frac{1}{m_{h'}^2}\right)^2 f_N^2 \frac{m_\Psi^4 m_N^4}{v_D^2 v^2(m_\Psi + m_N)^2 \pi},
    \\
    \sigma_\text{SI} (V N \to V N)
    =& s_\theta^2 c_\theta^2 
    \left( \frac{1}{m_h^2} - \frac{1}{m_{h'}^2}\right)^2 f_N^2 \frac{m_V^4 m_N^4}{v_D^2 v^2 (m_\Psi + m_N)^2 \pi},
    \\
    \sigma_\text{SI} (H N \to H N)
    =& s_\theta^2 c_\theta^2 
    \left( \frac{1}{m_h^2} - \frac{1}{m_{h'}^2}\right)^2 f_N^2 \frac{m_H^4 m_N^4}{v_D^2 v^2 (m_\Psi + m_N)^2 \pi},
\end{align}
where 
\begin{align}
&f_N = \frac{2}{9} + \frac{7}{9} \sum_{q=u,d,s} f_q,\\  &f_u = 0.0110,\ f_d = 0.0273,\ f_s = 0.0447.
\end{align}
These $f_u$, $f_d$, and $f_s$ are the default values used in \texttt{micrOMEGAs}~\cite{1801.03509}.

\section{Dark matter phenomenology}\label{sec:pheno}

In our setup, several particles are in the nontrivial representations of $T'$.
As we have discussed, $V$ and $H$ are $\mathbf{3}$, and $\Psi$ is $\mathbf{2}'+\mathbf{2}''$.
All of the other particles, including the SM particles, are singlets of $T'$.  
These nonsinglet particles cannot decay into the singlet particles, and 
thus the lightest particle among $V$, $H$, and $\Psi$ is a dark matter candidate in our setup. In addition, since $\Psi$ does not mix with SM fermions, fermion number conservation would imply that $\Psi$ is stable whether or not it is the lightest $T'$ nonsinglet.
As a result, depending on the mass spectra, the model contains potentially two dark matter particles.
If $2 m_{\Psi} < m_{V}, m_{H}$, then $V$ and $H$ can decay into $\Psi$, and thus $\Psi$ is the only dark matter candidate.
If $m_{V} < 2m_{\Psi} < m_{H}$, then $\Psi$ and $V$ are stable and both become DM particles.
If $m_{H} < 2m_{\Psi} < m_{V}$, then $\Psi$ and $H$ are DM particles.

We use \texttt{micrOMEGAs}~\cite{1801.03509} to calculate the DM relic abundance and the WIMP-nucleon scattering cross section.
The WIMP-nucleon scattering is induced by $h/h'$ exchange in the $t$-channel,
and the spin-independent cross section ($\sigma_{\text{SI}}$) is proportional to $\sin^2(2\theta)$. 
As we shall see, small $\theta$ is required to avoid the constraint from the Xenon1T experiment~\cite{1805.12562}.

\subsection{Single component fermionic DM}\label{sec:only-fermion-DM}
\begin{figure}[tb]
\includegraphics[width=0.3\hsize]{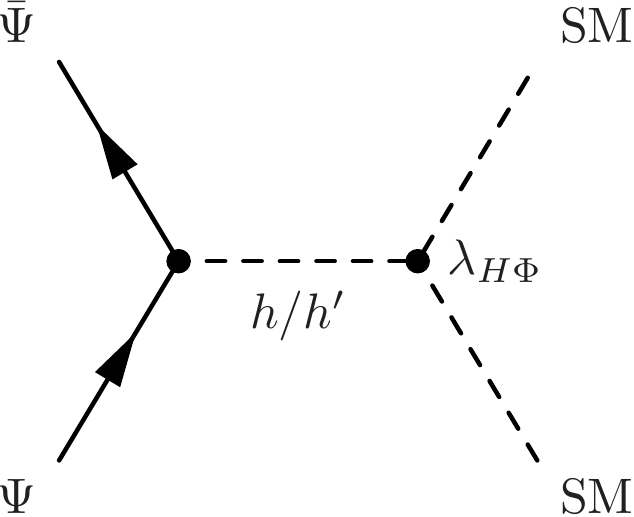} \\ \ \\
\includegraphics[width=0.3\hsize]{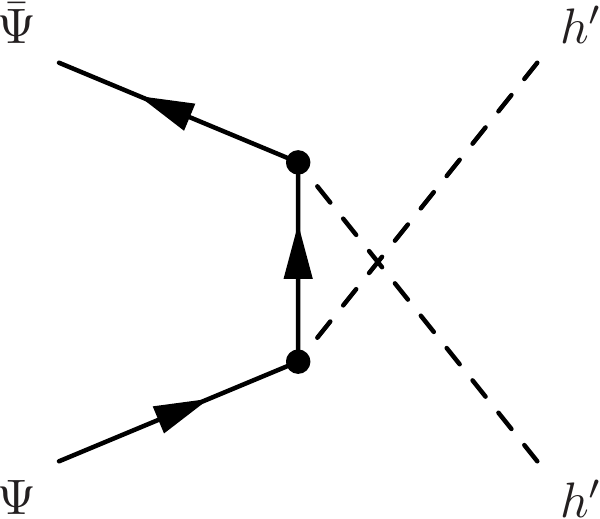} \quad
\includegraphics[width=0.3\hsize]{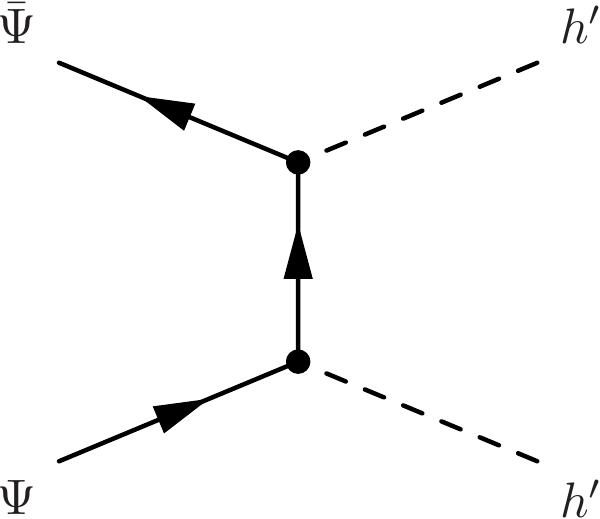} \quad
\includegraphics[width=0.3\hsize]{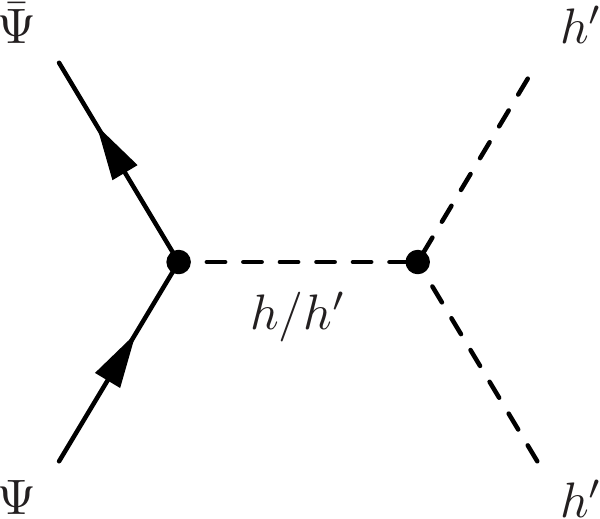} 
\caption{
Some annihilation channels for the fermionic DM particles. 
The diagram in the upper panel is the dominant annihilation channel for $m_{\Psi} \lesssim m_{h'}$.
The three diagrams in the lower panels show the fermionic DM annihilation process for $m_{\Psi} \gtrsim m_{h'}$.
}
\label{fig:ann}
\end{figure}

\begin{figure}[tb]
\includegraphics[width=0.8\hsize]{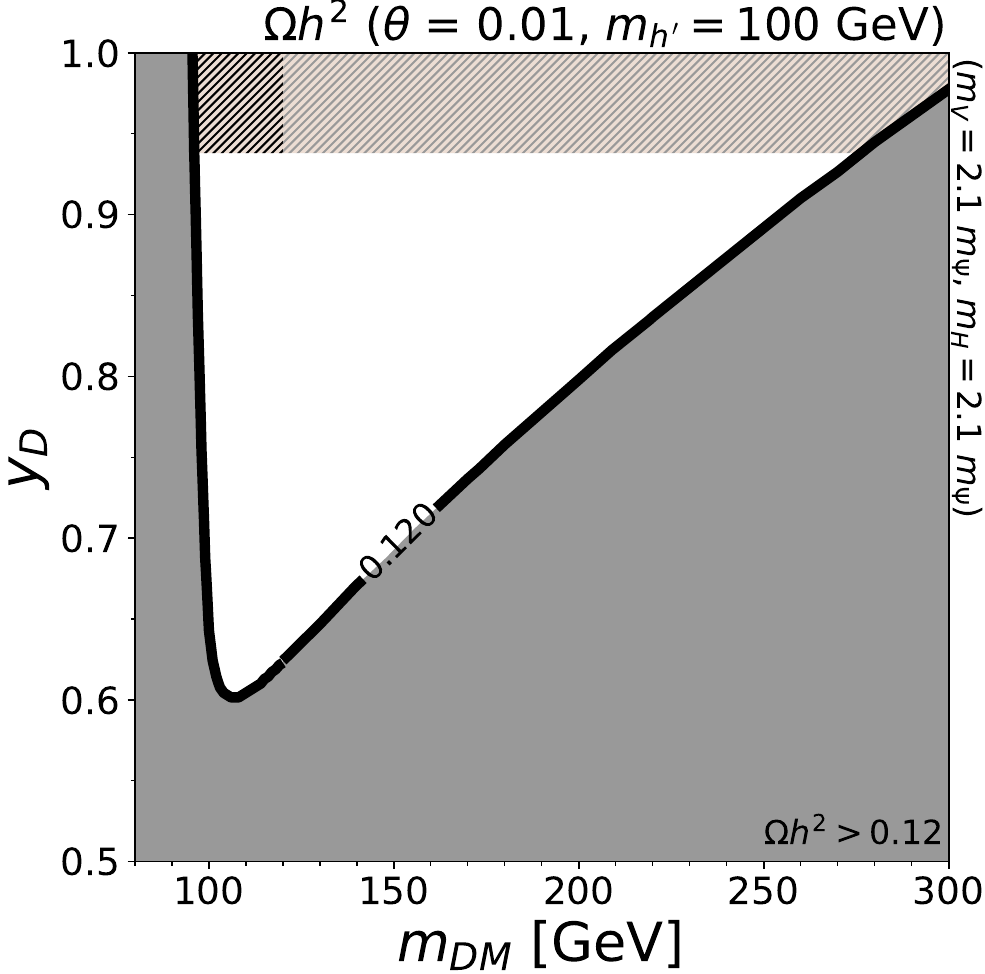} 
\caption{
Contours show values of $\Omega h^2$ in the $m_\Psi$-$y_D$ plane.
The red shaded regions with the hatched pattern is excluded by the PU bound on $g_D$.
In the gray-shaded region, dark matter is overabundant.
}
\label{fig:fermionic-DM-h'h'-1}
\end{figure}

We first consider the case where the DM candidate is a single component fermion.  This scenario is realized when  $2 m_\text{DM} < m_{H}, m_{V}$.
First, we investigate the case when
\begin{align}
  m_{h'} < 2 m_{\Psi} < m_{H,V},
\label{eq:mpsi-vs-mh'}
\end{align}
where the upper bound on $m_\Psi$ is determined to allow the triplet fields to decay into $\Psi$.
For this case, a DM pair can annihilate into $hh'$ and $h'h'$. 
Since the $\Psi$-$h'$ coupling is not suppressed by $\theta$, 
the dominant annihilation channel is $\Psi \Psi \to h'h'$.
The annihilation processes are shown in Fig.~\ref{fig:ann}.
Therefore, the model in this parameter regime can be classified into the secluded DM~\cite{0711.4866}.
This annihilation process is essential to obtain the right amount of DM energy density.
Figure~\ref{fig:fermionic-DM-h'h'-1} shows
 $\Omega h^2$ for $m_{h'} = 100$~GeV, $\theta=0.01$, $m_V=2.1 m_\Psi$, and $m_H = 2.1 m_\Psi$.
The measured value of the DM energy density is $\Omega h^2 = 0.1198\pm0.0012$~\cite{Aghanim:2018eyx}.
We find that the right amount of the DM relic density is obtained in a wide range of DM mass.
This is because the pair annihilation process of DM into $h'h'$ is efficient for $m_\Psi \gtrsim m_{h'}$.
As can be seen, around $m_\Psi =$ 100~GeV, the contour drastically changes. This is the sign of the opening of the $h'h'$ channel. 
Before the opening, the main annihilation channel is $b\overline{b}$, which is suppressed by small $\theta$,  and thus it requires a large $y_D$ for $\Omega h^2 = 0.12$. Once the $h'h'$ channel opens, there is no small $\theta$ suppression in the channel, and thus large $y_D$ is not required.
Since pairs of DM particles in the initial state have momentum, so the $h'h'$ channel opens slightly below 100~GeV. 
We also show the constraint from  DM direct detection by the Xenon1T experiment. 
We estimate the constraints from the Xenon1T experiment as follows.
In the direct detection experiments, the DM-nucleon scattering ratio is proportional to the DM number density
times the DM-nucleon scattering cross section, $n_\text{DM} \sigma_{\text{SI}}$. 
The number density is equal to the energy density divided by the DM mass, 
and the energy density is proportional to $\Omega h^2$. Therefore, we estimate the constraint from the 
direct detection experiment for the single DM particle scenario as
\begin{align}
 \sigma_{\text{SI}} \frac{\Omega h^2}{\Omega h^2_\text{obs.}} < \sigma_\text{exp.},
\end{align}
where $\Omega h^2_\text{obs.} \simeq 0.12$
and $\sigma_\text{exp.}$ is the upper bound given by the Xenon1T experiment.\footnote{
This scaling is expected in the case of cold DM~\cite{1005.4280},  
and is also used in the literature~\cite{1306.4710, 1709.01945,1906.06864}.
}
The constraint is stronger for larger $\theta$ because larger $\theta$ makes the DM-Higgs coupling larger. 
Note that the WIMP-nucleon scattering cross section is proportional to $\sin^2 (2\theta)$.
We have found that large values of $y_D$ are disfavored by the PU bound.
Note that $y_D$ is related to the gauge coupling and the scalar quartic couplings by Eq.~\eqref{eq:gD} and Eqs.~\eqref{eq:lam1}--\eqref{eq:lamHPhi}. 
We need to discuss the PU bound shown in Sec.~\ref{sec:PU},
because larger $y_D$ values lead to larger $\lambda_2$ and $g_D$; see Eqs.~\eqref{eq:gD} and \eqref{eq:lam2}.
We find that the PU bound on $g_D$ gives the strongest upper bound on $y_D$.
Both the PU bound and $\Omega h^2 \simeq 0.12$ are satisfied if $m_{h'} \lesssim 280$~GeV.

\begin{figure}[tb]
\includegraphics[width=0.48\hsize]{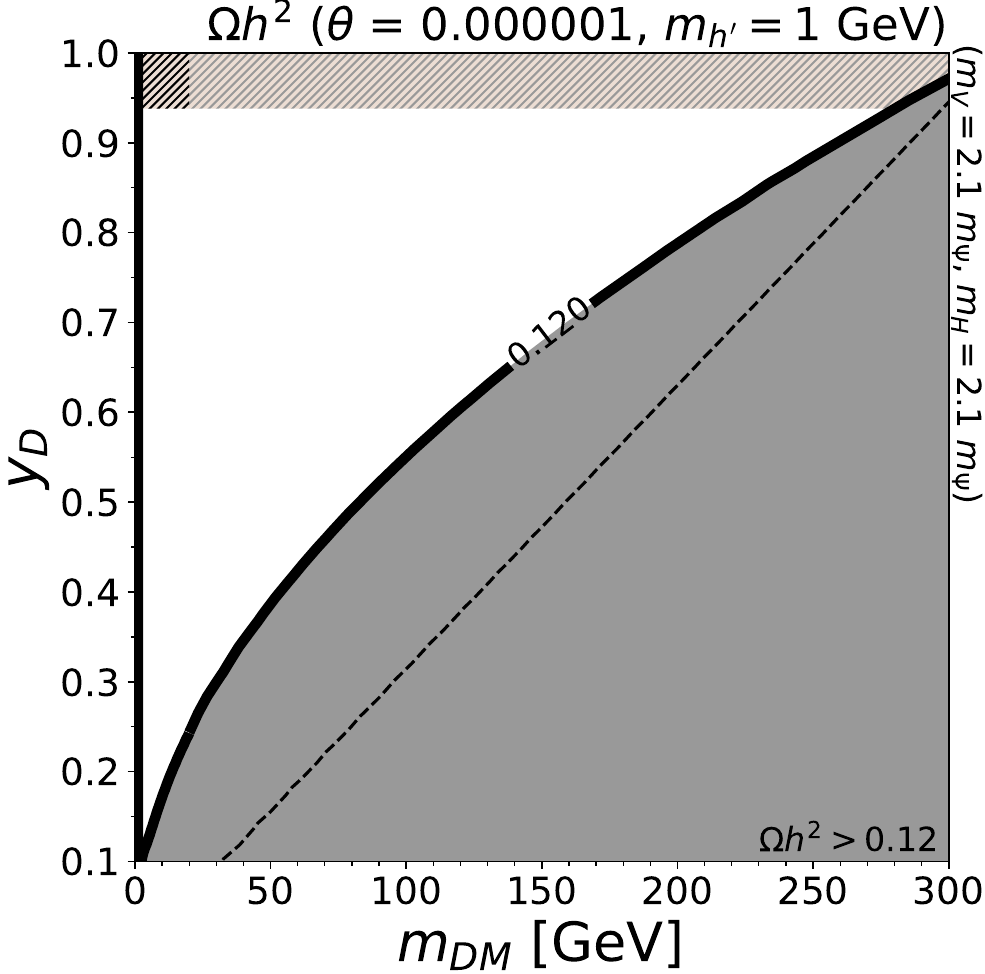} 
\includegraphics[width=0.48\hsize]{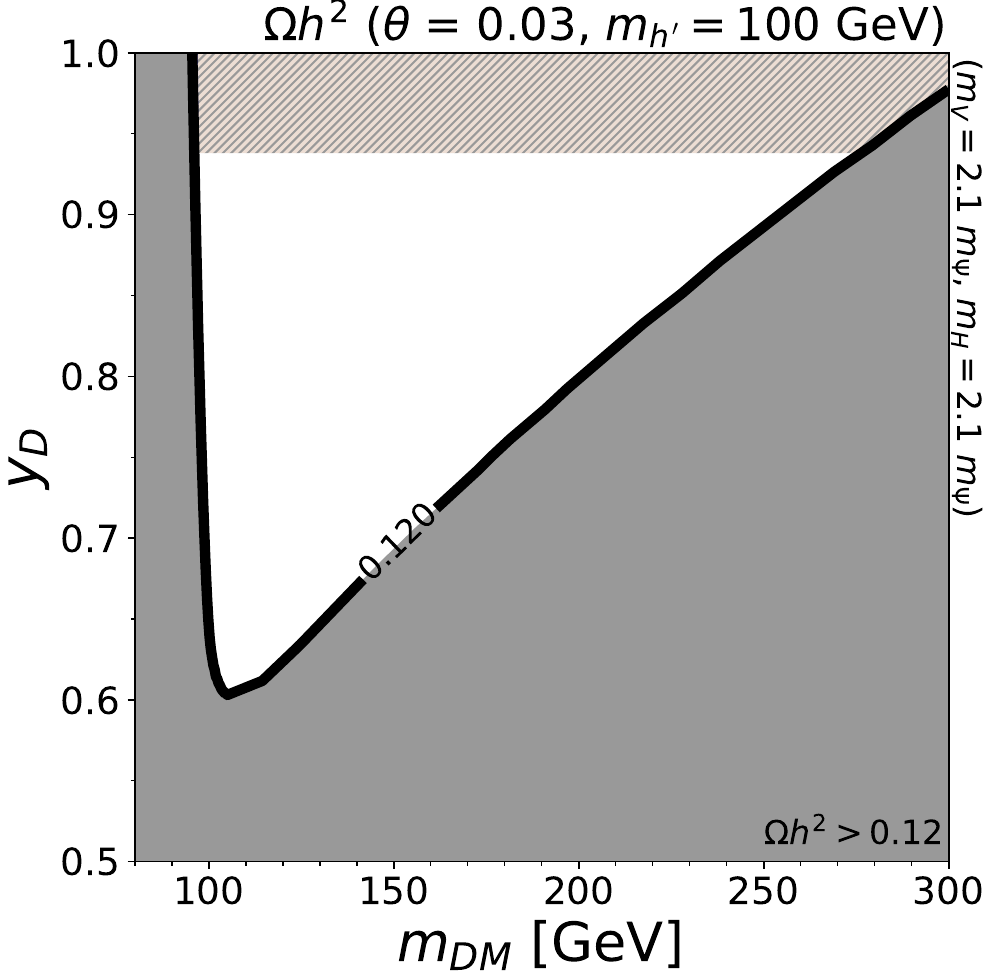} 
\includegraphics[width=0.48\hsize]{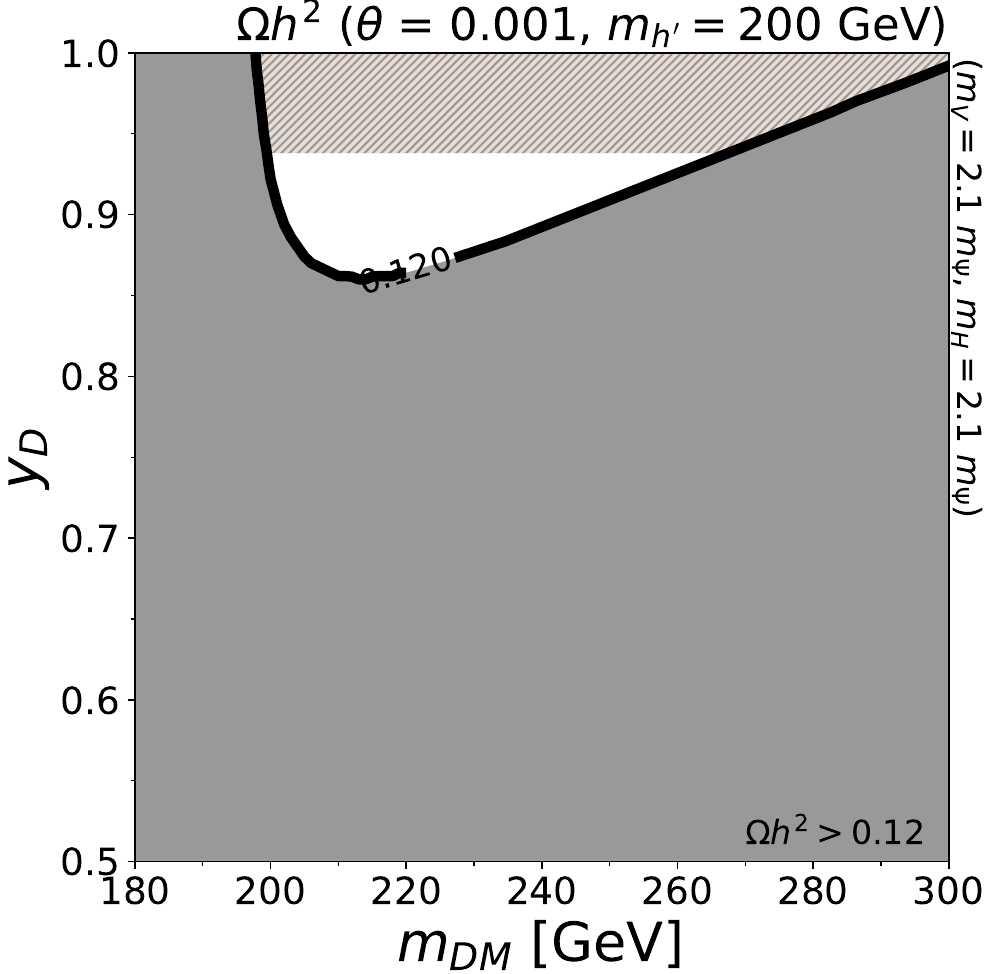} 
\includegraphics[width=0.48\hsize]{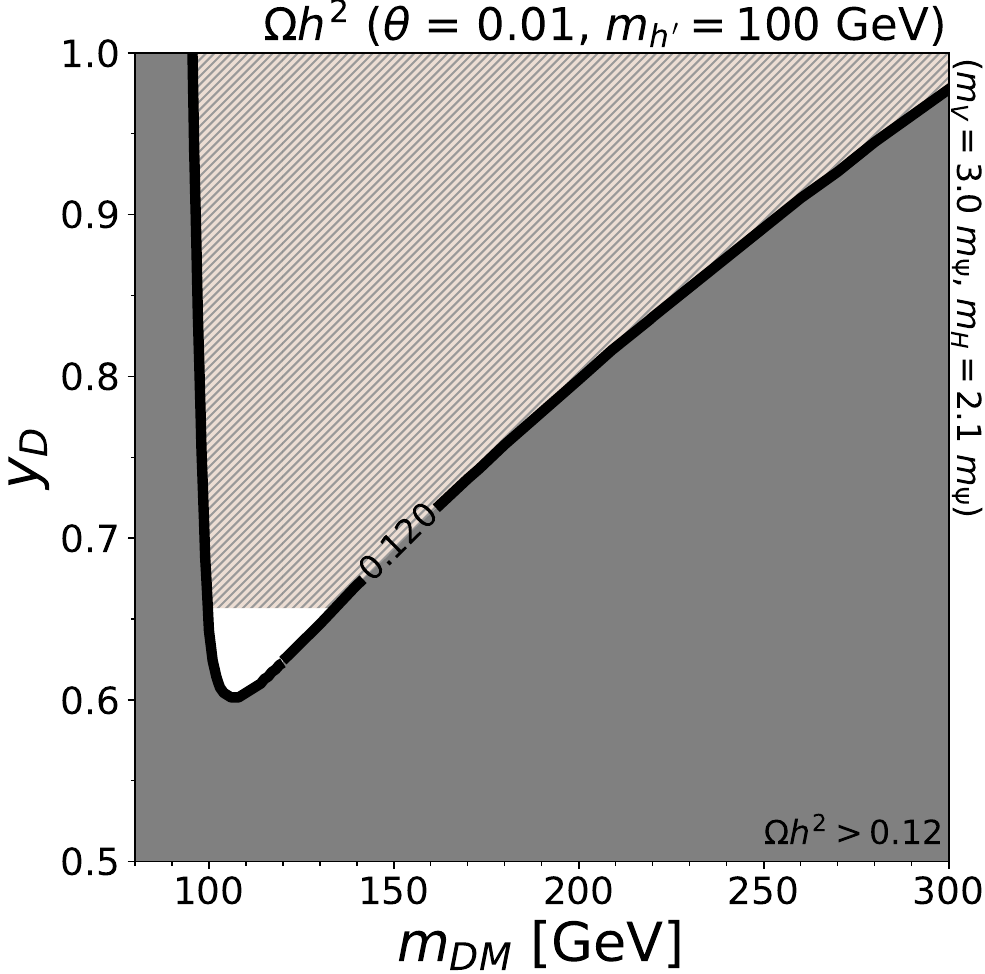} 
\caption{
Same as in Fig.~\ref{fig:fermionic-DM-h'h'-1} but for different parameter sets.
Along the dashed straight line, $\lambda_{H\phi} = 10^{-7}$. 
We cannot expect that $h'$ is in thermal equilibrium with the SM particles on the right side of the dashed straight line.
}
\label{fig:fermionic-DM-h'h'-2}
\end{figure}

Figure~\ref{fig:fermionic-DM-h'h'-2} is similar to Fig.~\ref{fig:fermionic-DM-h'h'-1} but with different parameter sets.
We find an upper bound on the DM mass, $m_{\text DM} \lesssim 280$~GeV, which does not strongly depend on the choice of $\theta$ or $m_{h'}$.
It does depend on the choice of $m_V$.
For larger $m_V$, the PU bound on $g_D$ gives the stronger constraint, as can be seen in the bottom-right panel of Fig.~\ref{fig:fermionic-DM-h'h'-2}, where $m_{V} = 3 m_\Psi$.
The contours for $\Omega h^2$ are not sensitive to the choice of $m_{V}$, but the PU bound is sensitive to it. 
We find $m_\text{DM} \lesssim 130$~GeV for $m_V = 3 m_\Psi$. Therefore, $m_V$ cannot be much heavier than the fermionic DM mass.

A smallish $\theta$ is required for a lighter $h'$ to avoid the constraint from the Xenon1T experiment. 
However, $h'$ and $\Psi$ do not thermalize with the plasma if the mixing angle $\theta$ is too small. 
There must be a lower bound on $\theta$ for the thermal relic scenario. 
We estimate it by comparing the interaction rate $\Gamma = n_{h'} \sigma_{h'h' \to hh}$ and the Hubble constant in the early Universe and find that $|\lambda_{H\phi}|$ should be larger than ${\cal O}(10^{-7})$. 
In the top-left panel of Fig.~\ref{fig:fermionic-DM-h'h'-2}, 
$|\lambda_{H\phi}| < 10^{-7}$ is on the right side of the dashed straight line.
The curve for the correct relic abundance is on the left side of the dashed straight line.
In the other panels, we find that $|\lambda_{H \phi}|$ is much larger than ${\cal O}(10^{-7})$.
Therefore, we can assume that $h'$ and $\Psi$ are in the thermal equilibrium in most of the parameter space.

For $m_{\text{DM}} < 100$~GeV, the indirect detection experiments might give constraints.
The dominant annihilation process of the DM pairs is $\Psi \bar{\Psi} \to h'h'$. 
This annihilation process is $p$-wave; see for example Ref.~\cite{1703.07364}.
Therefore, $\VEV{\sigma v}_{\Psi \bar{\Psi} \to h'h'}$ in the current Universe is smaller than the upper bound from experiments. We calculate it using \texttt{micrOMEGAs} and find that
$\VEV{\sigma v} \lesssim 10^{-29}$~cm$^3$/s. This is much less than the bound from the Fermi-LAT Collaboration~\cite{Fermi-LAT:2016uux}, ${\cal O}(10^{-26})$~cm$^3$/s. 
Therefore, there is no constraint from the indirect detection experiments.

\begin{figure}[tb]
\includegraphics[width=0.48\hsize]{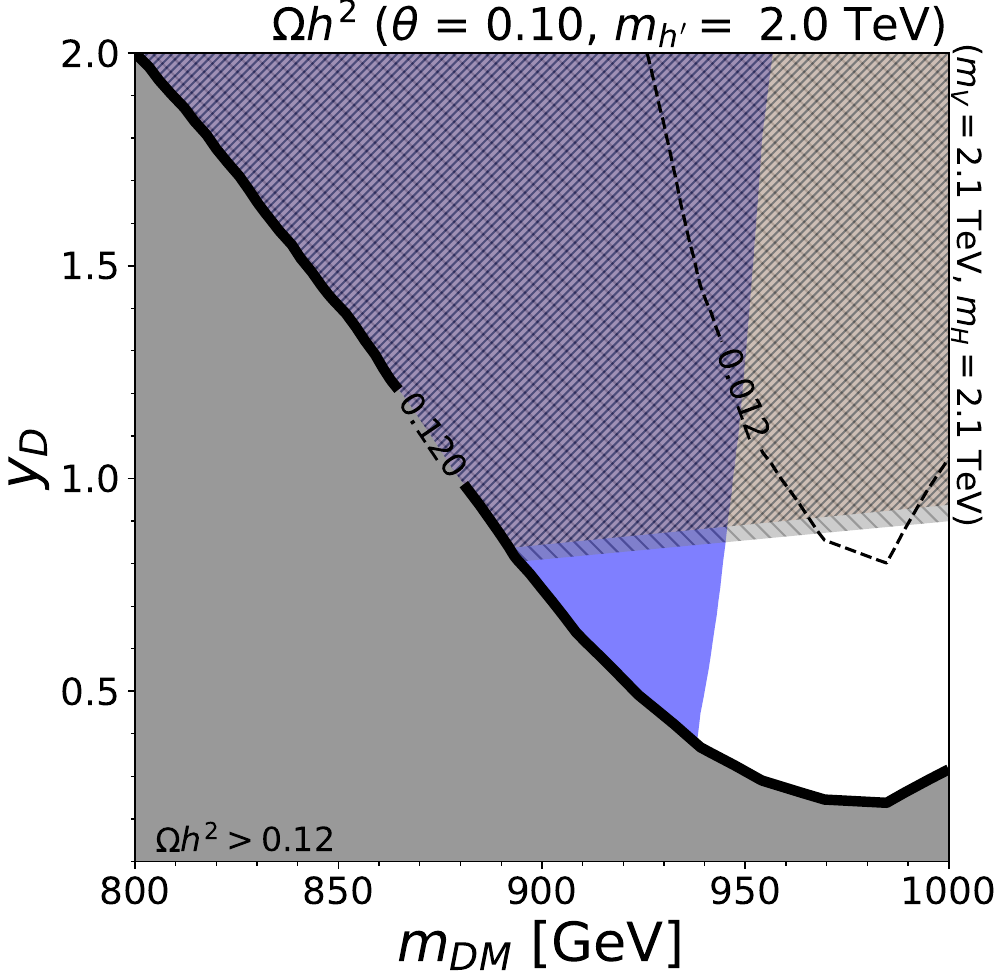} 
\includegraphics[width=0.48\hsize]{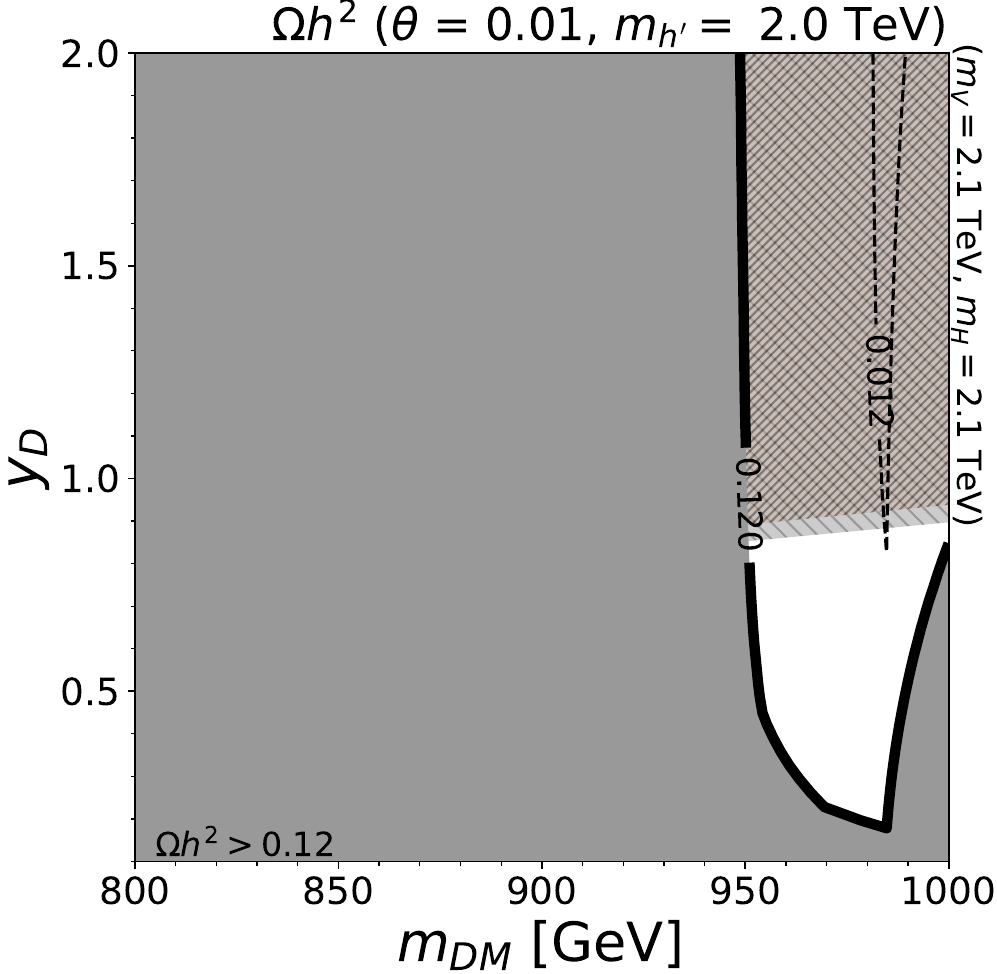} 
\caption{
Contours show values of $\Omega h^2$ in the $m_\Psi$-$y_D$ plane.
The lower gray shaded regions with the hatched pattern (\textbackslash \textbackslash) are excluded by the PU bound on $\lambda_1$ and $\lambda_2$.
The upper red shaded regions with the hatched pattern (//) are excluded by the PU bound on $g_D$.
}
\label{fig:MSS0_2000}
\end{figure}

Next, we consider the case where 
$2 m_{\text{DM}} \lesssim m_{h'}$.
In this case, the DM pairs annihilate only into the SM particles through
the $s$-channel exchange of $h$ and $h'$; see the diagram in the upper panel of Fig.~\ref{fig:ann}. 
We show the relic abundance of the fermionic DM in Fig.~\ref{fig:MSS0_2000}.
Here we take $m_{h'} = 2$~TeV, and $m_H = m_V = 2.1$~TeV.
Since all of the annihilation processes are suppressed by $\theta$, the annihilation cross section cannot be 
large enough to obtain the right amount of DM energy density.
We find that DM is overabundant in most regions of the parameter space.
An exception is for $m_{\text{DM}} \simeq m_{h'}/2$, 
where a DM pair annihilates efficiently thanks to a resonance of $h'$ in the $s$-channel. 
We also show the constraint from the DM direct detection by the XENON1T experiment. 
As can be seen from the figure, the XENON1T experiment gives the strong bound.
We conclude that we need to rely on the $h'$ resonance to obtain the correct amount of the relic abundance as long as the DM pairs annihilate only into the SM particles.

\begin{figure}[tb]
\includegraphics[width=0.48\hsize]{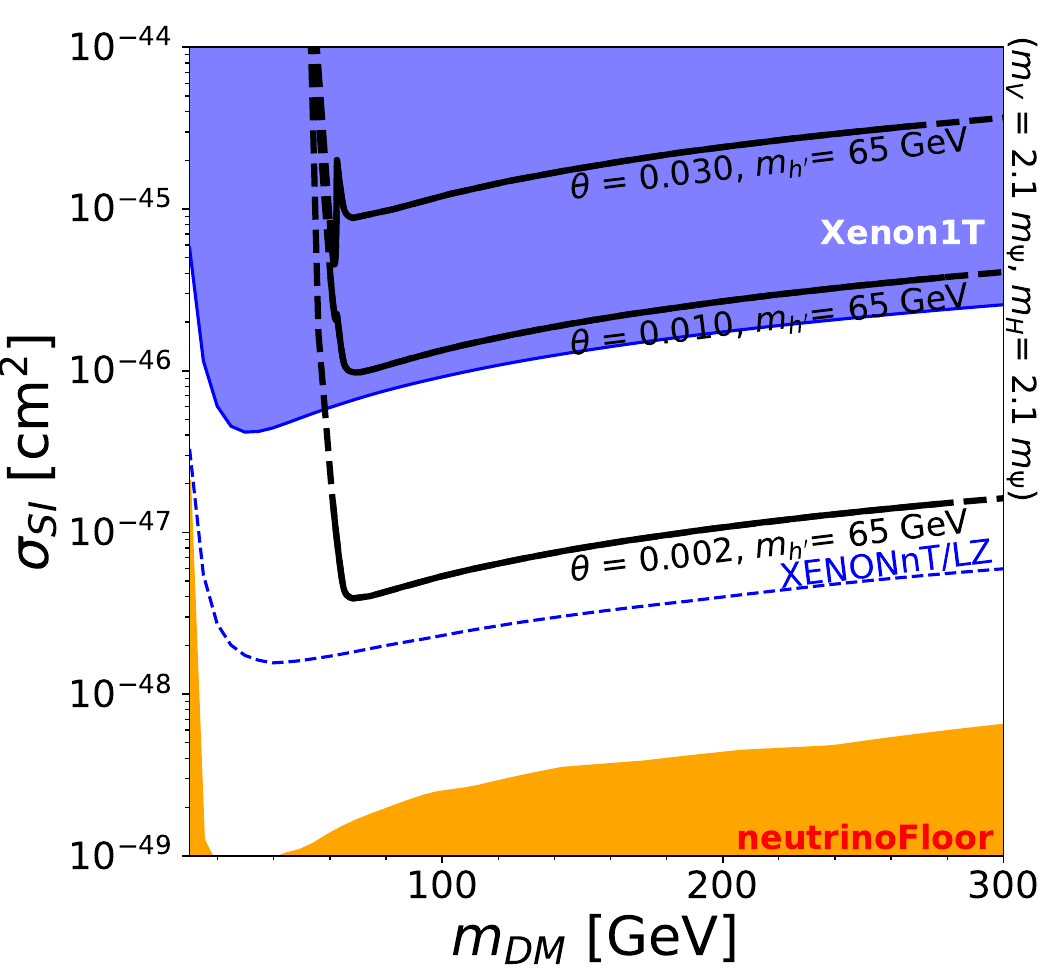} 
\includegraphics[width=0.48\hsize]{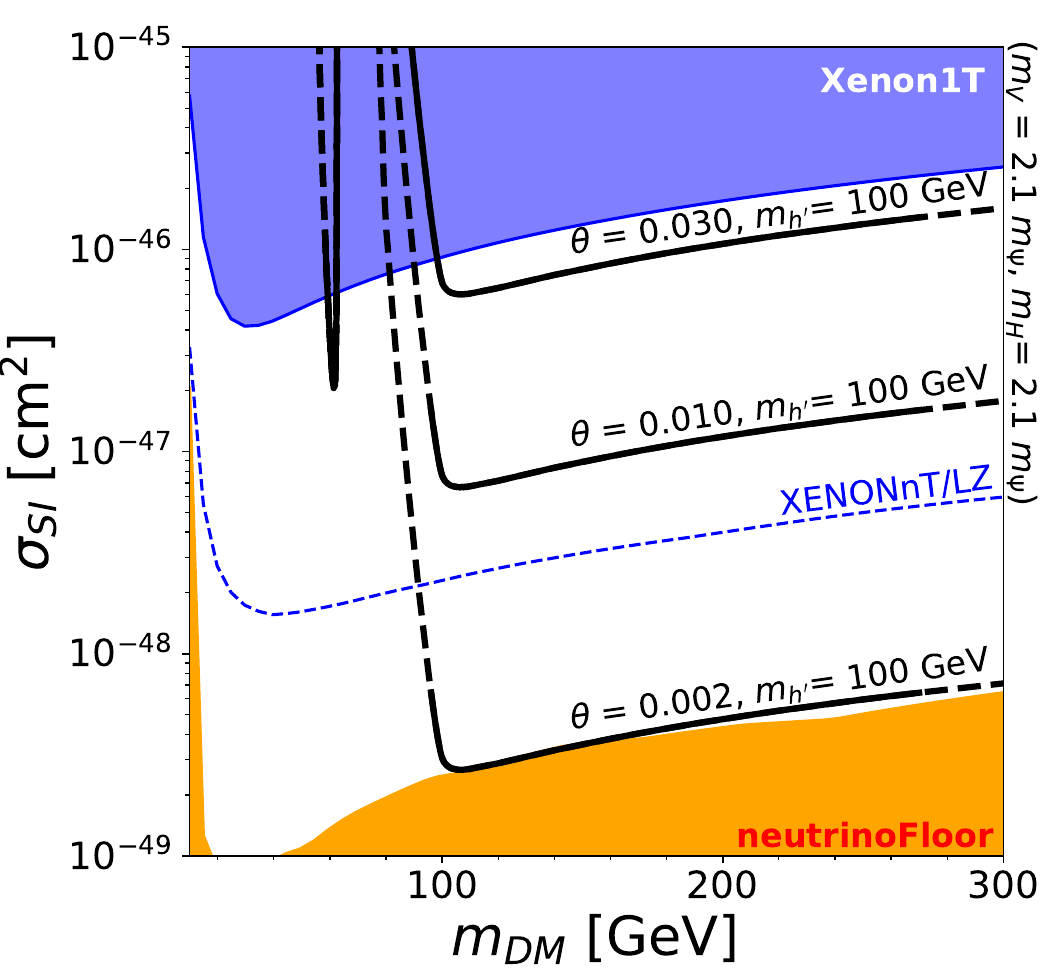} 
\includegraphics[width=0.48\hsize]{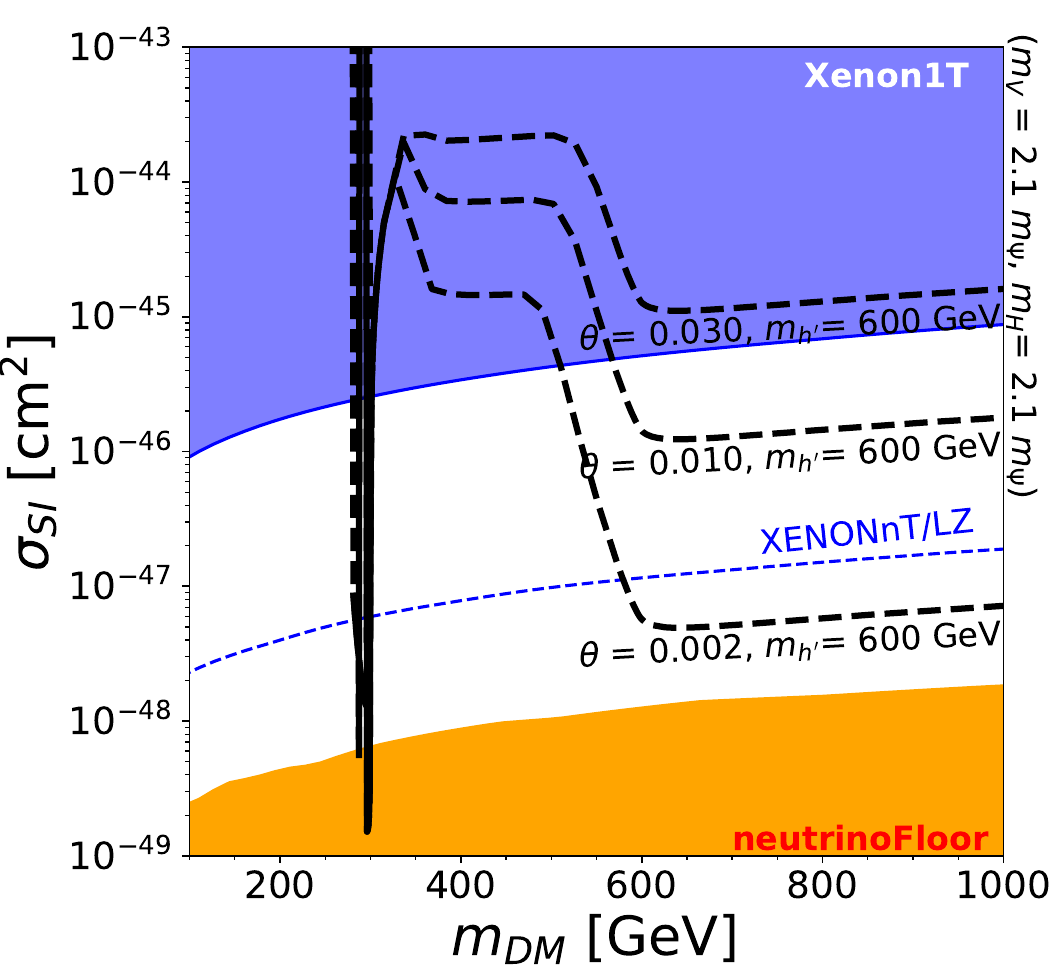} 
\caption{
Spin-independent cross section for $m_{h'}=65$, 100, and 600~GeV.
Here we choose $y_D$ to obtain the measured value of the DM energy density, $\Omega h^2 \simeq$ 0.12.
On the black dashed lines, the PU bound on $\lambda_1$ and $\lambda_2$ is not satisfied.
}
\label{fig:fermionic-DM_mass-vs-sigmaSI}
\end{figure}

Figure~\ref{fig:fermionic-DM_mass-vs-sigmaSI} shows the $\Psi$-nucleon spin-independent scattering cross section 
as a function of the DM mass for a given parameter set.
Here we fix $y_D$ to obtain the measured value of the DM energy density.
We find that $\theta$ should be small to avoid the constraint from the Xenon1T experiment.
This is because the $\Psi$-nucleon scattering cross section is induced by the $h$ and $h'$ exchange diagrams
and thus is proportional to $\sin^2 2\theta$.
For $\theta \gtrsim {\cal O}(10^{-3})$, 
it is possible to test this model in ongoing and future  direct detection experiments, such as 
Xenon1T~\cite{1512.07501}, LZ ~\cite{1802.06039}, 
and the DARWIN project~\cite{1606.07001}, which plan to reach down to the neutrino floor.
In the upper-left panel, three curves overlap with each other at $m_{\text{DM}} \simeq m_h/2$.
However, we find that they do not satisfy the PU bound for $\theta< 0.01$.
In that panel, the region for $m_{\text{DM}} < 57$~GeV is excluded by the constraint on the Higgs invisible decay~\cite{ATLAS:2018doi}.
For $m_{h'} \gtrsim 200$~GeV, we have to rely on the funnel region ($m_{\text{DM}} \simeq m_{h'}/2$) to obtain the measured value of the DM energy density.

It may be worthwhile to compare DM in this theory with other simple dark matter models, such as
the scalar-singlet DM model~\cite{Silveira:1985rk, McDonald:1993ex, Burgess:2000yq}, 
the inert doublet DM model~\cite{Barbieri:2006dq, Deshpande:1977rw}, 
and the singlet-doublet DM model~\cite{Mahbubani:2005pt, DEramo:2007anh, Enberg:2007rp, Cohen:2011ec, Cheung:2013dua}.
Apart from the resonant region, our theory is viable only for $m_{\text{DM}} < 280$~GeV. 
For this mass range, the singlet-scalar DM model and the inert doublet DM model are viable only at the Higgs funnel region, $m_{\text{DM}} \simeq m_h/2$~\cite{Abe:2014gua}.
The singlet-doublet DM model is viable for this mass range by using the blind spot~\cite{Cheung:2013dua} or the CP-violating dark sector~\cite{Abe:2014gua, Abe:2017glm, Abe:2019wku}. 
In the latter case, electric dipole moments are predicted, which are a feature different from our scenario.

\subsection{Very light single-component DM}

Here we briefly discuss the low-mass regime of DM. A lower bound $m_{DM} > 10$ GeV on the DM mass may be derived from a detailed analysis of the CMB data~\cite{Leane:2018kjk}
under the assumption that 
pairs of DM particles annihilate into two SM particles via an $s$-wave process. If the DM  pair annihilation results in beyond-SM particles, or if the process is suppressed by $p$ wave, then this lower bound is not applicable, and much lighter DM is allowed, with masses as low as MeV. We briefly discuss this case here. We have carried out the DM abundance analysis in the case of a single-component fermionic DM with mass below 10 GeV.

\begin{figure}[tb]
\includegraphics[width=0.48\hsize]{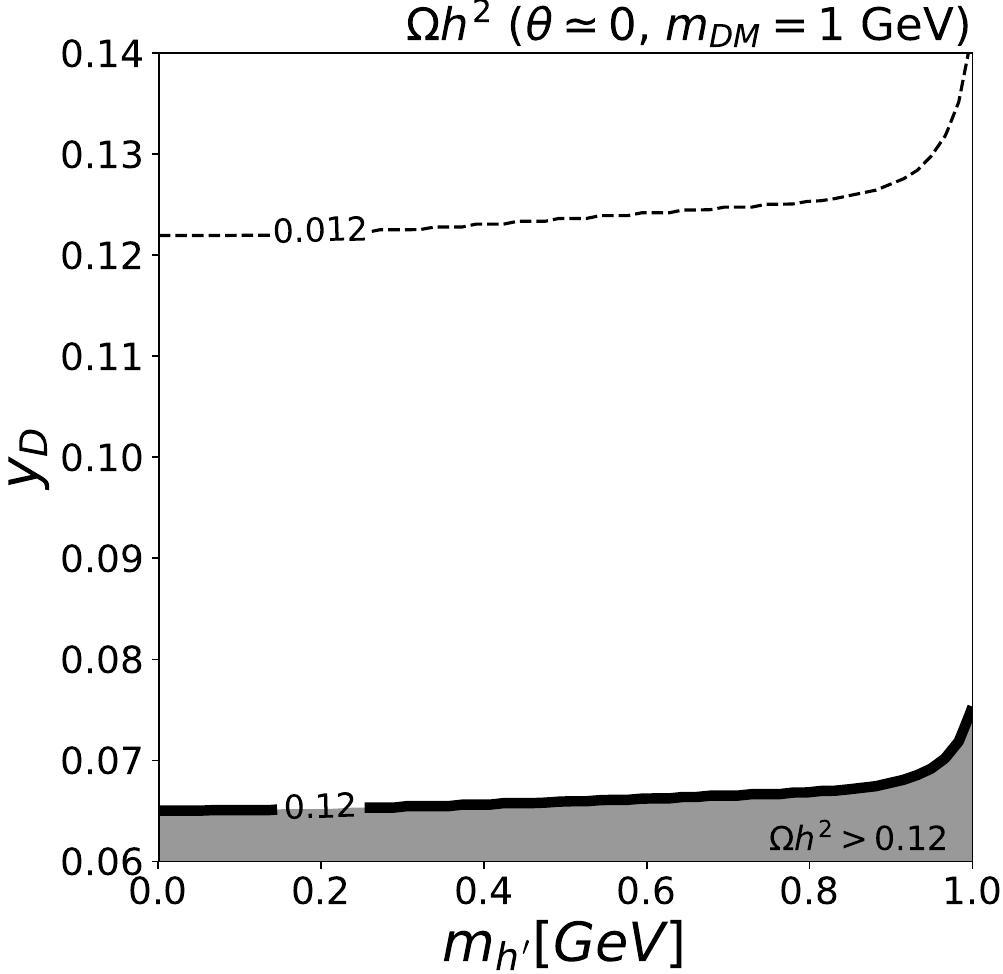} 
\includegraphics[width=0.48\hsize]{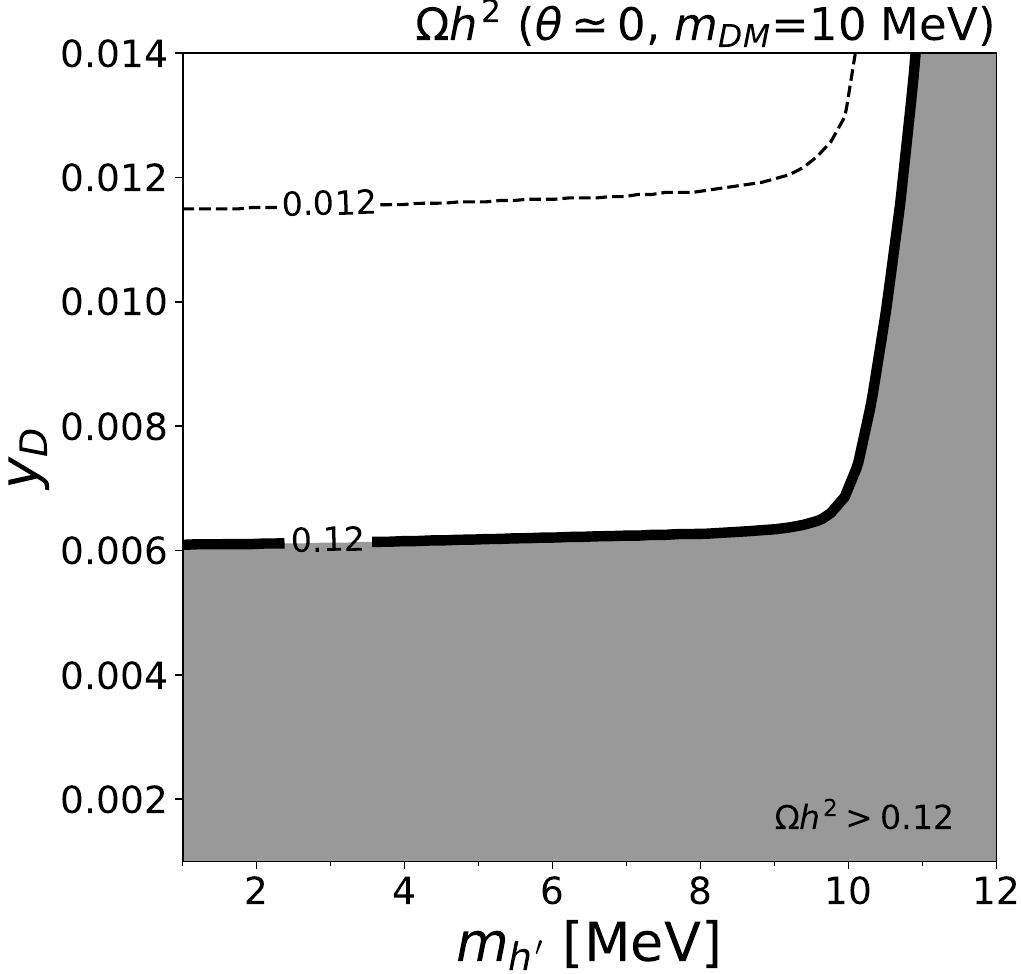} 
\caption{
The contours show the value of $\Omega h^2$ in $m_{h'}$-$y_D$ plane.
}
\label{fig:Omegah2-lightDM}
\end{figure}

As can be seen from the top-left panel of Fig.~\ref{fig:fermionic-DM-h'h'-2}, if pairs of DM particles annihilate into pairs of $h'$,
then we can obtain the right amount of the relic abundance and also keep $y_D$ in the perturbative regime. Thus, we focus on $m_{h'} \lesssim m_\text{DM}$. Figure~\ref{fig:Omegah2-lightDM} shows the contours of $\Omega h^2$ in the $m_{h'}$-$y_D$ plane. We take $m_\text{DM} = 1$~GeV (10~MeV) in the left (right) panel. We take a very small $\theta$, and the result is independent of its precise value as long as $\theta \ll {\cal O}(10^{-3})$. The right amount of relic abundance is always obtained for $m_{h'} < m_\text{DM}$. 

The lower mass bound on $m_\text{DM}$ is estimated from the lower mass bound on $h'$. Thus, we investigate the lower bound on the mass of $h'$. For $m_{h'} < 1$~GeV, $h'$ can decay into $\mu \bar{\mu}$, $e^- e^+$, and $\gamma \gamma$. The decay width is proportional to $\sin^2\theta$, and thus the lifetime of $h'$ is longer for smaller values of $\theta$. If the lifetime of $h'$ is longer than $\sim 1$~sec., then it may affect the big bang nucleosynthesis. We require its lifetime to be shorter than 1 sec 
and then find that $\theta > 4.8 \times 10^{-8}$ is required for $m_{h'}$ = 250~MeV, where $h'$ mainly decays into $\mu \bar{\mu}$.
As another example, we find that $\theta > 5.5 \times 10^{-5}$ is required for $m_{h'}$ = 2~MeV, where $h'$ mainly decays into $e^- e^+$.
A more dedicated study of the constraints on $\theta$ and $m_{h'}$ has been done in Ref.~\cite{Krnjaic:2015mbs}, from which we find that the lower bound on $m_{h'}$ is about 1~MeV. 
Combining this result and the result in the previous paragraph, 
we conclude that $m_\text{DM}$ can be as low as $\sim 1$~MeV with the freeze-out mechanism.


\subsection{Two-component DM scenario}
\label{sec:3+Psi}
Next, we consider the case where $H$ or $V$ cannot decay into $\Psi$, 
namely, $m_\mathbf{3} < 2 m_\Psi$, where $m_\mathbf{3} = \min(m_V, m_H)$.
In this case, the lightest $T'$ triplet is also a DM candidate in addition to $\Psi$.
The total DM energy density is the sum of the energy density of the doublet and triplet fields, $\Omega h^2 = \Omega h^2_\Psi + \Omega h^2_\mathbf{3}$, where $\Omega h^2_\mathbf{3}$ is for the lighter one of $H$ and $V$.

We can divide this case into two cases:
$m_\mathbf{3} < m_\Psi$ and $m_\Psi < m_\mathbf{3} < 2 m_\Psi$.
In the former case, fermionic DM pairs can annihilate into the triplets,
and thus 
it is expected that the dominant component of the DM is the triplet.
In the latter case, $\mathbf{3} \mathbf{3} \to \Psi \Psi$ is expected, and thus $\Psi$ is expected to be the dominant component of DM.
If $m_\Psi \simeq m_{h'}/2$, 
then the phenomenology for DM physics is similar to the case discussed in Sec.~\ref{sec:only-fermion-DM}, and thus we do not discuss the case with $m_\Psi \simeq m_{h'}/2$ here.

In a similar manner as the single-component DM scenario, we estimate the constraint from 
direct-detection experiments for multiple DM particles as
\begin{align}
 \sum_i \frac{\Omega h^2_i}{m_i} \sigma_{\text{SI}}{}_i  < \frac{\Omega h^2_\text{obs.}}{m_\text{DM}} \sigma_\text{exp.}
\end{align}
where $\Omega h^2_\text{obs.} \simeq 0.12$, $m_\text{DM}$ is the DM mass assumed in the analysis by the experiments,
and $\sigma_\text{exp.}$ is the upper bound given by the experiment for $m_\text{DM}$.
If there is only one species of dark matter, this equation gives the standard relation, 
$ \sigma_{\text{SI}} < \sigma_\text{exp.}$.
At first glance, it is unclear how to choose $m_\text{DM}$ in our analysis because we have more than one species of DM.
However, for $m_\text{DM} \gtrsim 200$~GeV, $\frac{\sigma_\text{exp.}}{m_\text{DM}}$ is almost constant, 
and thus we can freely choose $m_\text{DM}$ if $m_i > 200$~GeV. In the following analysis, we choose 
\begin{align}
 m_\text{DM} = \frac{\Omega h^2_\Psi m_\Psi + \Omega h^2_\mathbf{3} m_\mathbf{3}}{\Omega h^2_\Psi + \Omega h^2_\mathbf{3}}.
\end{align}

\begin{figure}[tb]
\includegraphics[width=0.48\hsize]{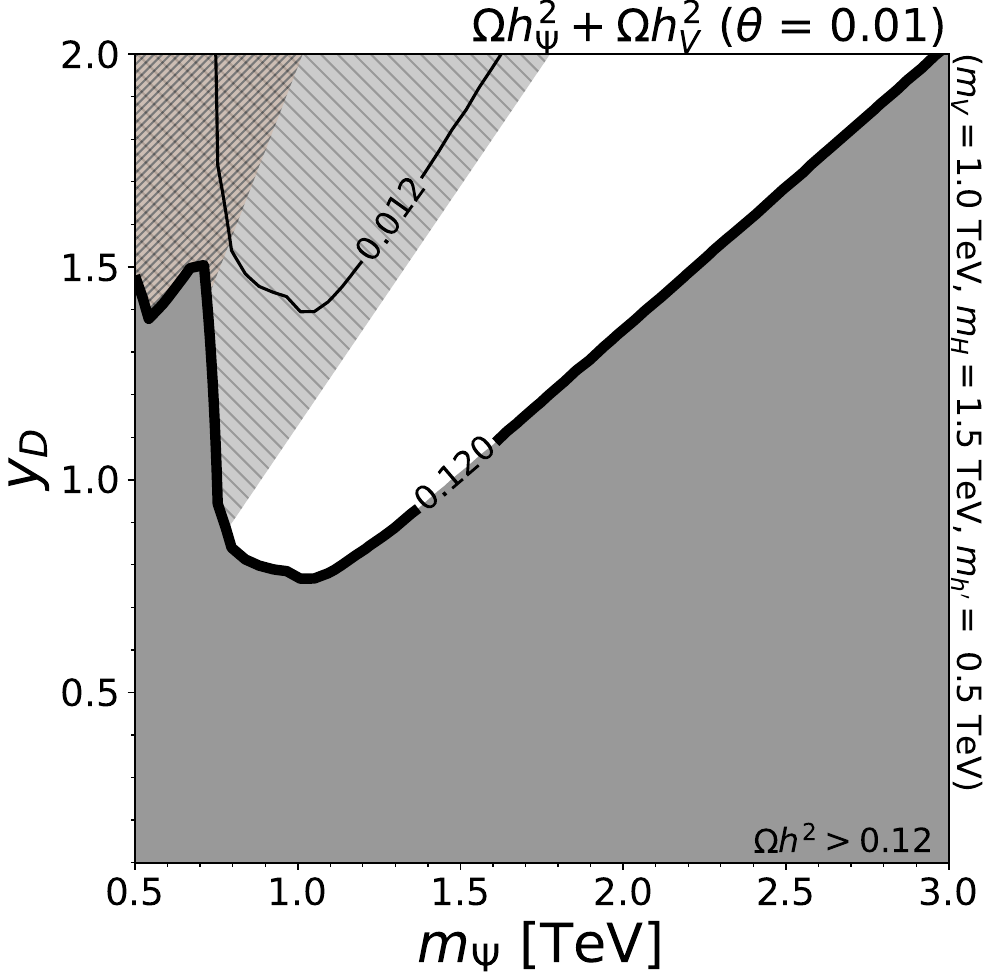} 
\includegraphics[width=0.48\hsize]{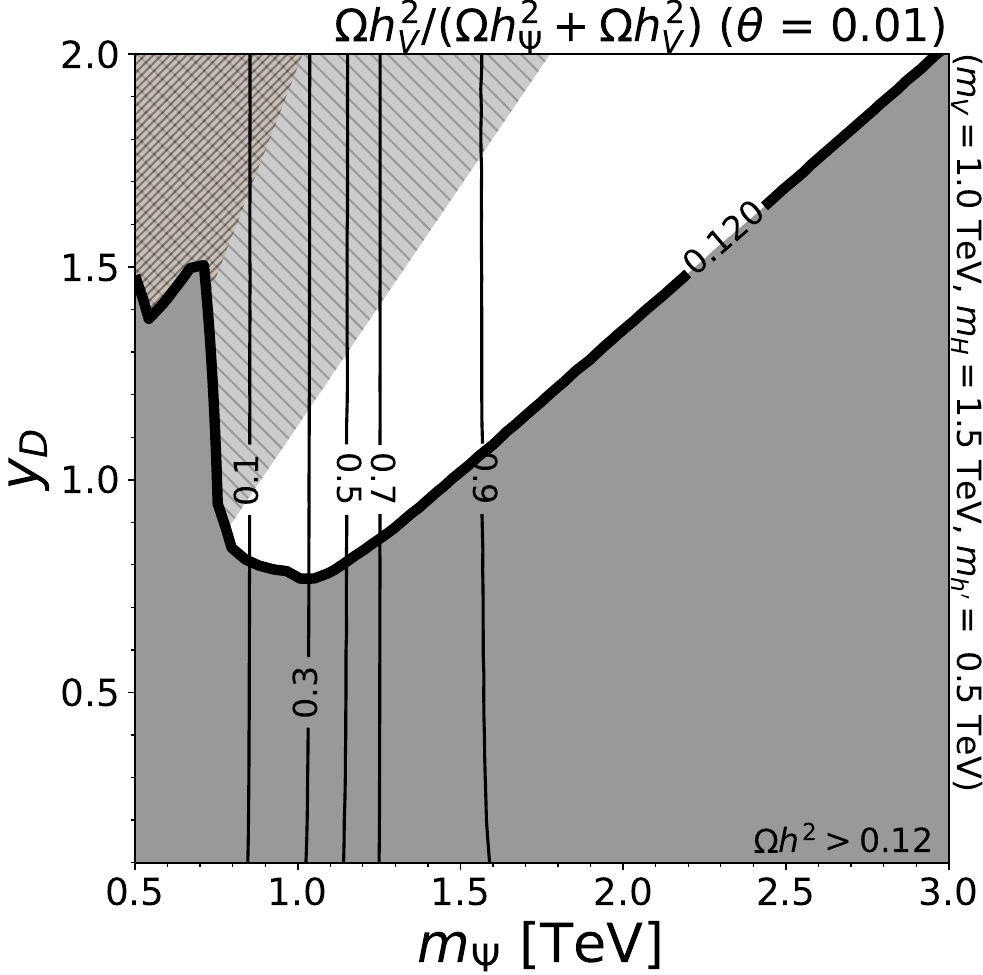} 
\caption{
\textbf{Left:} contours show the values of $\Omega h^2_\Psi + \Omega h^2_V$ in the $m_\Psi$-$y_D$ plane. The color notations are the same as in Fig.~\ref{fig:MSS0_2000}. 
\textbf{Right:} proportion of the vector DM.
}
\label{fig:tripletDM-v1}
\end{figure}

Figure~\ref{fig:tripletDM-v1} shows $\Omega h^2$ contours for the vector DM + fermionic DM.
Here we take $m_V = 1$~TeV, $m_H= 1.5$~TeV,  $m_{h'} =0.5$~TeV, and $\theta = 0.01$ as a benchmark.
We find that the vector DM is the dominant component (more than 50\% of the DM energy is occupied by the vector DM) for $m_\Psi \gtrsim 1.1$~TeV. See the right panel in Fig.~\ref{fig:tripletDM-v1}. 
In the vector DM dominant region, the PU bound is weak. 
This is a different feature of this case from the single-component fermionic DM case discussed in Sec.~\ref{sec:only-fermion-DM}.
For $m_\Psi < 750$~GeV, the main annihilation process for the fermionic DM is $\Psi \Psi \to h' h'$.
The vector DM pairs annihilate into $h'h'$ and $\Psi \Psi$.
Since $h'$ is lighter than $\Psi$, $VV \to h'h'$ is the main process.
Both the $\Psi \Psi \to h' h'$ and $VV \to h' h'$ annihilation cross sections are proportional to $y_D^4$.
For $m_\Psi > 750$~GeV, the fermionic DM can annihilate into $Vh'$. The annihilation cross section of this process is proportional to $g_D^2 y_D^2$, and thus a smaller $y_D$ is sufficient to obtain the measured value of the DM energy density. This behavior can be seen in the left panel of Fig.~\ref{fig:tripletDM-v1}.
For $m_\Psi > 1$~TeV, the $\Psi \Psi \to H h'$ and $\Psi \Psi \to VV$ channels are open. The latter channel does not depend on $y_D$, and thus smaller $y_D$ is required as can be seen in the left panel of Fig.~\ref{fig:tripletDM-v1}.
For $m_\Psi > 1$~TeV, the vector DM is the dominant component. 
The main annihilation channel is $VV \to h'h'$, and the annihilation cross section is proportional to $g_D^4$.
Since $g_D$ is proportional to $y_D/m_\Psi$, see Eq.~\eqref{eq:gD},
a larger $y_D$ is required for the larger $m_\Psi$ regime. 

\begin{figure}[tb]
\includegraphics[width=0.48\hsize]{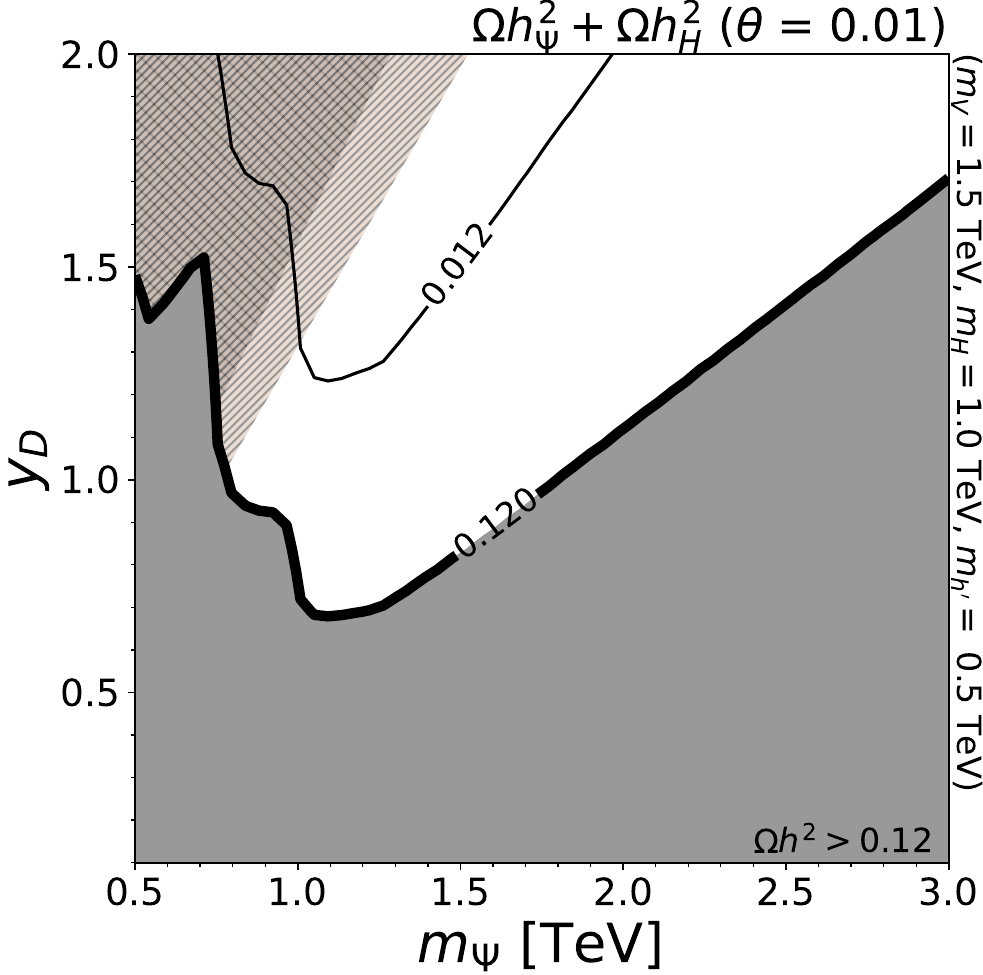} 
\includegraphics[width=0.48\hsize]{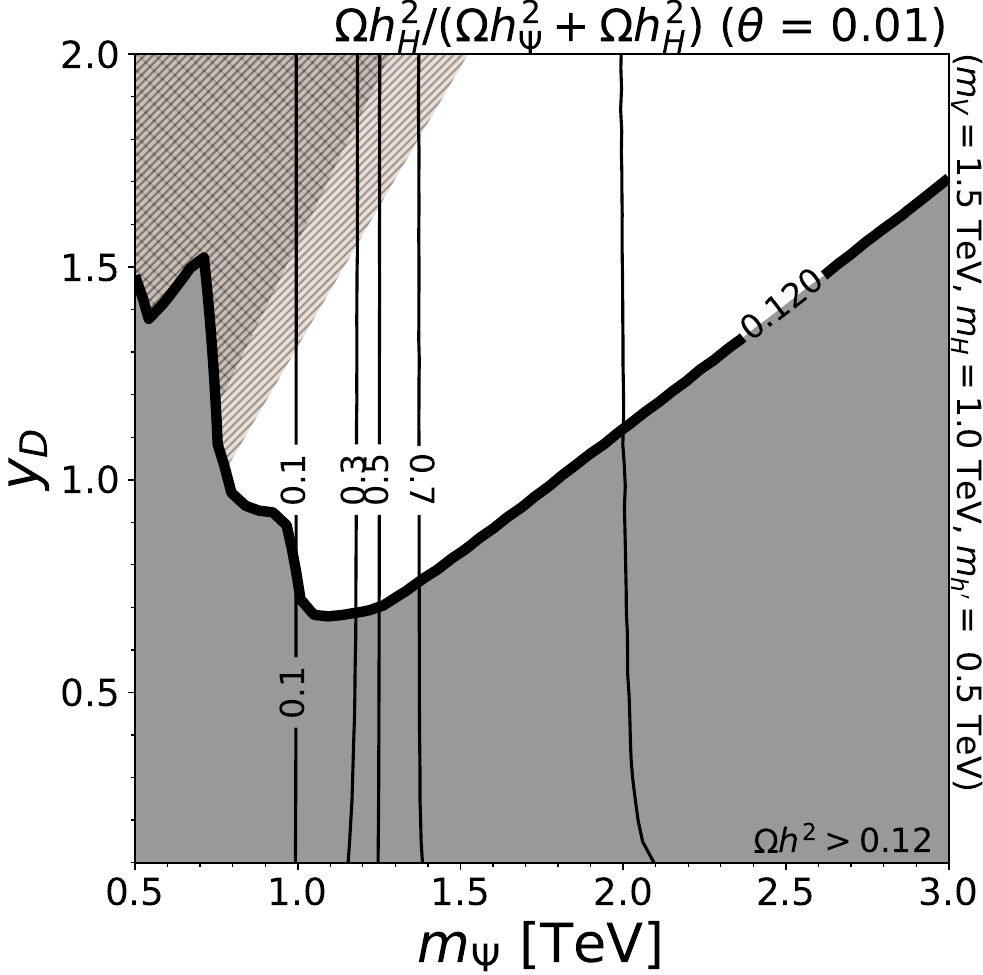} 
\caption{
\textbf{Left:} contours show the values of $\Omega h^2_\Psi + \Omega h^2_H$ in the $m_\Psi$-$y_D$ plane. The color notations are the same as in Fig.~\ref{fig:MSS0_2000}. 
\textbf{Right:} proportion of the scalar DM.
}
\label{fig:tripletDM-scalar}
\end{figure}

Figure~\ref{fig:tripletDM-scalar} shows $\Omega h^2$ contours for the scalar DM + fermionic DM.
Here we take $m_V = 1.5$~TeV, $m_H= 1$~TeV,  $m_{h'} =0.5$~TeV, and $\theta = 0.01$ as a benchmark.
The result is similar to the vector DM + fermionic DM case, but a smaller $y_D$ is sufficient to obtain the measured value of the DM energy density.

As a byproduct of the current work, we have found a model with a discrete
non-Abelian gauge symmetry. If we do not introduce the 4-plet fermion
into the theory, a single-component bosonic DM scenario can be realized.  The stability of such a DM candidate is guaranteed by the unbroken $T'$ (in this case actually an $A_4$) symmetry.  We have focused here on chiral fermionic DM, motivated by the understanding it provides regarding the DM mass scale. Purely bosonic DM scenarios are also interesting, and some of their phenomenology can be inferred from the present analysis.

\section{Conclusions}\label{sec:conclusion}

In this paper, we have proposed a simple theory of chiral fermionic dark matter.  It is motivated by a desire to explain the scale of the DM mass as well as its stability.  Thermal DM should have a mass in the range ${\cal O}(10)$~GeV--${\cal O}(100)$~TeV in order to explain the DM energy density by the freeze-out mechanism.  Most DM models take the mass scale as an input.  The chiral DM theory presented here links the DM mass to the scale of a dark sector $SU(2)_D$ gauge symmetry breaking.  
The chiral fermion is the 4-plet representation under the gauge symmetry. This theory is perhaps the simplest chiral theory that is free of anomalies.  A scalar 7-plet field is introduced to achieve spontaneous symmetry breaking and generate mass for the chiral fermion DM. After the scalar field develops a VEV, a discrete non-Abelian symmetry $T'$ remains unbroken. Consequently, the theory admits a two-component DM scenario, with one component being the chiral fermion, and the other being the lightest of $T'$ nonsinglet boson.  

We have studied the internal consistency requirements of the theory.  Perturbative unitarity constrains the gauge coupling $g_D$, the Yukawa coupling $y_D$, and the quartic scalar couplings. In the single-component fermionic DM scenario,
we found that the measured value of the DM energy density is explained for $m_\Psi \lesssim 280$~GeV and $m_{h'} \lesssim 200$~GeV, or for $m_\Psi \simeq m_{h'}/2$. To avoid the constraint from the Xenon1T experiment, the Higgs mixing parameter $\theta$ should be small. If $\theta \gtrsim {\cal O}(10^{-3})$, we have a chance to see DM signals at DM direct experiments in the near future.
We have also investigated the two-component DM scenario, where both bosonic and fermionic DM particles exist. In this case, we found a larger region of viable parameter space.

\section*{Acknowledgments}
This work is supported in part by JSPS KAKENHI Grant Numbers 16K17715 and 19H04615 (T.A) and by the US Department of Energy
Grant Number DE-SC 0016013 (K.S.B). T.A. wishes to thank the High Energy Physics Group at Oklahoma State University for its hospitality during his visit in May 2019 where part of the work was done. K.S.B. acknowledges the Kobayashi-Maskawa Institute for warm hospitality during a visit where this work was initiated.  K.S.B wishes to thank Bogdan Dobrescu and Andre de Gouvea for helpful discussions.  He also thanks the Fermilab Theory Group for warm hospitality during a summer visit under the Fermilab Distinguished Visitor program.

\appendix
\section*{Appendix}

\section{Transformation rules in $T'$}
\label{sec:T'rules}
There are 24 group elements in $T'$, and four of them are generators~\cite{1003.3552},
$e$, $s$, $t$, and $r$, where $e$ is an identity element. They satisfy
\begin{align}
 s^2 = r, \ r^2 =t^3  = e, \ rt = tr.
\end{align}
Under $r$, $s$ and $t$, the triplet fields transform as
\begin{align}
 r: \quad &
 X^1 \to X^1, \
 X^2 \to X^2, \
 X^3 \to X^3, \\ 
 s: \quad &
 X^1 \to X^1, \
 X^2 \to -X^2, \
 X^3 \to -X^3, \\ 
 t: \quad  &
 X^1 \to X^3, \
 X^2 \to X^1, \
 X^3 \to X^2,
\end{align}
where $X^j = \pi_j, H_j$, and $V^j_\mu$ in this model.
These transformations are the same as the transformation of triplets in $A_4$.
The transformation rules of $\mathbf{2'}$ and $\mathbf{2''}$ are
\begin{align}
 r:\quad & \psi_{\mathbf{2'}} \to -\psi_{\mathbf{2'}},\quad
    \psi_{\mathbf{2''}} \to -\psi_{\mathbf{2''}},\\
 s:\quad & \psi_{\mathbf{2'}} \to \begin{pmatrix}
			    0 &  i \\  i & 0
			   \end{pmatrix} \psi_{\mathbf{2'}},\quad
 \psi_{\mathbf{2''}} \to \begin{pmatrix}
			    0 &  i \\  i & 0
			   \end{pmatrix} \psi_{\mathbf{2''}},\\
 t:\quad &  \psi_{\mathbf{2'}} \to -\frac{\omega}{\sqrt{2}}\begin{pmatrix}
			    \bar{p} & -p  \\ \bar{p} & p
			   \end{pmatrix} \psi_{\mathbf{2'}},\quad
 \psi_{\mathbf{2''}} \to -\frac{\omega^2}{\sqrt{2}}\begin{pmatrix}
			    \bar{p} & -p  \\ \bar{p} & p
			   \end{pmatrix} \psi_{\mathbf{2''}},
\end{align}
where 
\begin{align}
 p = \exp\left( i \frac{\pi}{4}\right),\quad
\bar{p} = \exp\left( -i \frac{\pi}{4}\right),\quad
 \omega = \exp\left( i \frac{2\pi}{3}\right).
\end{align}

\section{Explicit expression of the scalar potential}
\label{sec:explicit-scalar-potential}
The quartic couplings in the scalar potential have the following expanded form when expressed in terms of mass eigenstates:

\begin{align}
 \left.V\right|_{\lambda_1, \lambda_2}=&
\quad 
\frac{1}{4}\lambda_1
\left( \vec{H} \cdot \vec{H}  + \vec{\pi} \cdot \vec{\pi} + \sigma_7^2 + 2 v_D \sigma_7 \right)^2
\nonumber\\
&
+
\lambda_2
\Biggl\{
\frac{5}{2}
\left(
\sqrt{\frac{3}{5}} H_1^2  - \sqrt{\frac{3}{5}} H_2^2  
+ H_1 \pi_1 + H_2 \pi_2 -2 H_3 \pi_3
\right)^2
+
\nonumber\\
&
\qquad
+
\frac{1}{2}
\left( H_1^2 + H_2^2 - 2 H_3^2 - \sqrt{15} H_1 \pi_1 + \sqrt{15} H_2 \pi_2 \right)^2
\nonumber\\
&
\qquad
+
\frac{5}{8}
\left(
 \sqrt{\frac{3}{5}} H_1 H_2 + H_1 \pi_2 - H_2 \pi_1 + 15 \pi_1 \pi_2 
+ 4 H_3 (v_D + \sigma_7)
\right)^2
\nonumber\\
&
\qquad
+
\frac{5}{8}
\left(
 \sqrt{\frac{3}{5}} H_2 H_3 + H_2 \pi_3 - H_3 \pi_2 + 15 \pi_2 \pi_3 
+ 4 H_1 (v_D + \sigma_7)
\right)^2
\nonumber\\
&
\qquad
+
\frac{5}{8}
\left(
 \sqrt{\frac{3}{5}} H_3 H_1 + H_3 \pi_1 - H_1 \pi_3 + 15 \pi_3 \pi_1 
+ 4 H_2 (v_D + \sigma_7)
\right)^2
\Biggr\}
,
\end{align}
where
\begin{align}
 \vec{X} \cdot \vec{X} = \sum_{j=1}^3 X_j X_j.
\end{align}

\section{Explicit expression of the scalar kinetic term}
The scalar kinetic term is written in terms of the component fields as
\begin{align}
& \frac{1}{2} \partial^{\mu} \vec{H} \cdot \partial_\mu \vec{H}
+ \frac{1}{2} \partial^{\mu} \vec{\pi} \cdot \partial_\mu \vec{\pi}
+ \frac{1}{2} \partial^{\mu} \sigma_7 \cdot \partial_\mu \sigma_7
+ \frac{1}{2} (4 g_D^2 v_D^2) \vec{V}^{\mu} \cdot \vec{V}_{\mu}
+ 2 g_D v_D \vec{V}^{\mu} \cdot \partial_\mu \vec{\pi}
\nonumber\\
&
+ 4 g_D^2 v_D \sigma_7 \vec{V}^{\mu} \cdot \vec{V}_{\mu}
+ 2 \sqrt{15} g_D^2 v_D \left( H_3 V^{1 \mu}V^2_\mu + H_2 V^{3 \mu}V^1_\mu + H_1 V^{2 \mu}V^3_\mu   \right)
\nonumber\\
&
+g_D^2
\Biggl\{
\frac{1}{2} 
\Biggl[
  6 H_1^2 (V^{2\mu} V^2_\mu + V^{3\mu}V^{3}_\mu )
+ 6 H_2^2 (V^{3\mu}V^{3}_\mu + V^{1\mu}V^1_\mu ) 
+ 6 H_3^2 (V^{1\mu}V^1_\mu + V^{2\mu}V^2_\mu) 
\nonumber\\
& \qquad \qquad
+ 3
\left(
  H_1 H_2 V^{1\mu} V^2_\mu
+ H_2 H_3 V^{2\mu} V^3_\mu 
+ H_3 H_1 V^{3\mu} V^1_\mu 
\right)
\nonumber\\
& \qquad \qquad
+ 4 (\vec{\pi} \cdot \vec{\pi} )(\vec{V}^{\mu} \cdot \vec{V}_\mu )
+ 15 
\left( V^{1\mu} V^2_\mu \pi_1 \pi_2 
+ V^{2\mu} V^3_\mu \pi_2 \pi_3 
+ V^{3\mu} V^1_\mu \pi_3 \pi_1
\right)
\Biggr]
\nonumber\\
& \qquad 
+ \frac{\sqrt{15}}{2}
 \Biggl[
  2 H_1  \pi_1 (V^{2\mu}V^2_\mu -  V^{3\mu} V^3_\mu)
+ H_1  V^{1\mu} (V^2_\mu \pi_2 -  V^{3}_{\mu} \pi_3) 
\nonumber\\
& \qquad  \qquad\qquad
+   
   2 H_2\pi_2 (V^3_\mu V^{3\mu}  -  V^{1\mu}V^1_\mu )
 + H_2 V^2_\mu (\pi_3 V^{3\mu}  - V^{1\mu} \pi_1)
\nonumber\\
& \qquad  \qquad \qquad
 + 2 H_3\pi_3 (V^{1\mu}V^1_\mu  - V^{2\mu} V^2_\mu)
 + H_3 V^{3\mu} (V^1_\mu  \pi_1 -  V^2_\mu \pi_2 )
\Biggr] 
\nonumber\\
& \qquad 
+ 2 \sqrt{15} 
\sigma_7 (H_1 V^{2\mu} V^3_\mu + H_2  V^{3\mu} V^1_\mu + H_3 V^{1\mu} V^2_\mu) 
+ 2 \sigma_7^2 (V^{1\mu}V^1_\mu + V^{2\mu}V^2_\mu + V^{3\mu}V^{3}_\mu) 
\Biggr\}
.
\end{align}

\section{Explicit expression of terms involving the Dirac fermion}
\label{app:fermion-int}
Using the Dirac fermions, terms in the Lagrangian with the fermion field can be written
with the Dirac fermion fields as 
\begin{align}
{\cal L}\supset 
& \bar{\Psi} \gamma^{\mu} \partial_\mu \Psi
 - y_D v_D \bar{\Psi} \Psi
 - y_D \sigma_7 \bar{\Psi} \Psi
\nonumber\\
&
+\frac{1}{2}g_D \bar{\Psi}
\begin{pmatrix}
 V^3_\mu &  V^1_\mu - i V^2_\mu \\
 V^1_\mu + i V^2_\mu & - V^3_\mu
\end{pmatrix}
\gamma^{\mu} 
\Psi
+g_D 
\left(
\bar{\Psi}
\begin{pmatrix}
 V^3_\mu &   \omega V^1_\mu - i \omega^2 V^2_\mu \\
 \omega V^1_\mu + i \omega^2 V^2_\mu & - V^3_\mu
\end{pmatrix}
\gamma^{\mu} P_L
\Psi^c
+
(h.c.)
\right)
\nonumber\\
&
-y_D
\left(
 \bar{\Psi}
\frac{i}{2}
\begin{pmatrix}
 \pi_V^3 &    \omega \pi_V^1 - i \omega^2 \pi_V^2 \\
 \omega \pi_V^1 + i \omega^2 \pi_V^2 & - \pi_V^3
\end{pmatrix}
\Psi^c
+(h.c.)
\right)
\nonumber\\
&
-\frac{2}{\sqrt{5}}y_D
 \bar{\Psi}
\begin{pmatrix}
 H_3 &  H_1 -i H_2 \\
 H_1 + i H_2 & - H_3
\end{pmatrix}
\Psi
-
\frac{y_D}{2 \sqrt{5}}
\left(
 \bar{\Psi}
\begin{pmatrix}
 H_3 & \omega H_1 - i \omega^2 H_2 \\
\omega H_1 + i \omega^2 H_2 & - H_3
\end{pmatrix}
\Psi^c
+(h.c.)
\right)
.
\end{align}

\end{document}